\documentclass{article}
\usepackage{arxiv}
\usepackage[utf8]{inputenc}
\usepackage[T1]{fontenc}
\usepackage{hyperref}     
\usepackage{booktabs}     
\usepackage{amsfonts}     
\usepackage{nicefrac}     
\usepackage{microtype}
\usepackage{lipsum}
\usepackage{bm}
\usepackage{amsmath}
\usepackage{multirow}
\usepackage{graphicx}
\graphicspath{ {./images/} }
\usepackage{subcaption}
\usepackage[authoryear,round]{natbib}
\usepackage{xcolor}
\usepackage{soul}
\RequirePackage[thmmarks]{ntheorem}

\title{Transformed-Linear Models for Time Series Extremes}

\author{
  Nehali Mhatre\\
  Department of Statistics\\
  Colorado State University\\
  \texttt{nmhatre@colostate.edu} \\
  
   \And
 Daniel Cooley \\
  Department of Statistics\\
  Colorado State University\\
  \texttt{cooleyd@stat.colostate.edu} \\

}

\theoremheaderfont{\hskip\parindent\normalfont\scshape}
\theorembodyfont{\itshape}
\theoremseparator{.}
\newtheorem{proposition}{Proposition}

\begin{document}
\maketitle

\begin{abstract}
In order to capture the dependence in the upper tail of a time series, we develop non-negative regularly-varying time series models that are constructed similarly to classical non-extreme ARMA models.
Rather than fully characterizing tail dependence of the time series, we define the concept of weak tail stationarity which allows us to describe a regularly-varying time series through the tail pairwise dependence function (TPDF) which is a measure of pairwise extremal dependencies.
We state consistency requirements among the finite-dimensional collections of the elements of a regularly-varying time series and show that the TPDF's value does not depend on the dimension being considered.
So that our models take nonnegative values, we use transformed-linear operations. 
We show existence and stationarity of these models, and develop their properties such as the model TPDF's.
Additionally, we show the class of transformed-linear MA($\infty$) models forms an inner product space.
Motivated by investigating conditions conducive to the spread of wildfires, we fit models to hourly windspeed data and find that the fitted transformed-linear models produce better estimates of upper tail quantities than traditional ARMA models or than classical linear regularly varying models.

\end{abstract}


\section{Introduction}
\label{sec:introduction}
Measuring risk associated with rare events that are extreme in magnitude requires extreme value methods.
Univariate extremes methods are well-developed, but there is a need to develop easily implementable statistical methods that can describe and model extremal dependence in the time series context.
This paper will use transformed-linear operations to construct straightforward and flexible models for nonnegative regularly-varying time series. 

\begin{figure}
\centering
    \includegraphics[width=0.8\textwidth, trim = 0cm 1.5cm 0cm 1cm, clip]{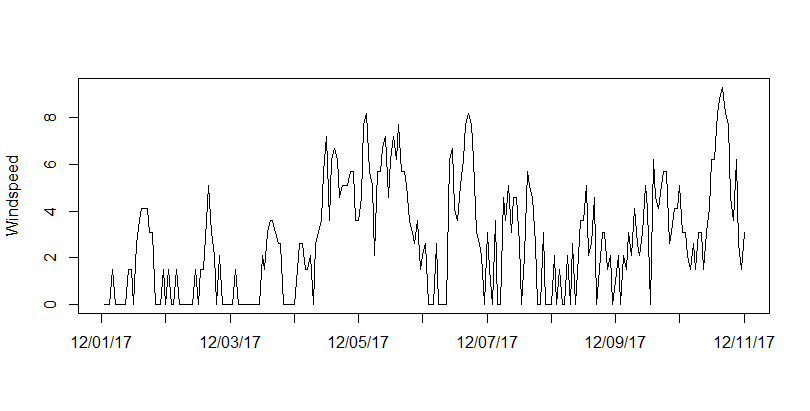}
    \includegraphics[width=0.8\textwidth, trim = 0cm 1.5cm 0cm 1.5cm, clip]{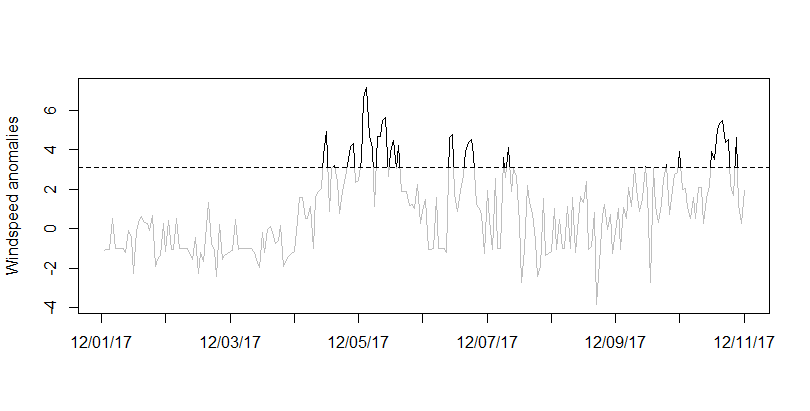}
    \caption{Time series of hourly windspeed (m/s) for 12/01/2017 through 12/11/2017. The Thomas fire started on 12/04/2017. Original windspeeds (upper panel) and windspeed anomalies (lower panel): bold above a threshold of 3.09 (95$\textsuperscript{th}$ percentile).}
    \label{Figure1}
\end{figure}

To motivate our models, consider the time series of hourly windspeeds (m/s) observed at the March AFB station in Southern California (HadISD \citep{dunn2019}; revisited in Section \ref{sec:santaana}).
High windspeeds are one of the factors that contribute to wildfire risk.
Figure \ref{Figure1} (upper panel) shows the hourly windspeeds for December 1-10, 2017.
The windspeeds increase for a period of time beginning on December 4, the ignition day of the Thomas Fire, a wildfire attributed to the weather phenomenon known as the Santa Ana winds.

Windspeed data are not stationary, as there is a diurnal cycle. To remove the diurnal behavior, we subtract off the mean from each hour of a 24-hour cycle creating a time series of windspeed anomalies that can reasonably be assumed stationary (Figure \ref{Figure1}, lower panel, data available \href{https://www.stat.colostate.edu/~cooleyd/TransLinTS/}{\textit{here}}).
Our goal is to model the upper tail behavior of this time series.
Exploratory analysis indicates the data at short lags exhibit asymptotic dependence. 
Two realizations $X_t$ and $X_{t+h}$ of a stationary time series with marginal distribution $F$ are said to be asymptotically independent if $\chi(h)=\text{lim}_{u\to1}P\{F(X_{t+h})>u|F(X_t)>u\}=0$, and asymptotically dependent if $\chi(h) \ne 0$ with the value of $\chi(h)$ summarizing the magnitude of pairwise tail dependence at lag $h$.
A chi-plot (not shown) for the upper tail shows that  $\hat \chi(1) \approx 0.55$.
Few probabilistic frameworks allow for asymptotic dependence.

Linear time series models, which include the familiar ARMA models, have many attractive qualities such as simplicity, interpretability, and widespread familiarity.
Much of classical time series analysis can be done with only an assumption of weak or second-order stationarity, which allows one to focus on the autocovariance function (ACVF).
This classical approach is poorly suited for describing extremal behavior as the ACVF describes dependence at the center of the distribution, and classical ARMA models are not designed with tail behavior in mind.

In this paper, we build time series models via transformed-linear combinations of regularly-varying terms.
The framework of regular variation is useful as it is defined only in terms of the tail, it is naturally linked to extreme value theory, and it is well-suited for modeling data that exhibit asymptotic dependence.
Models take only nonnegative values in order to focus attention on the upper tail.
To construct models that feel similar to linear models, we use transformed-linear operations \citep{cooleythibaud2019} which define addition and multiplication in a manner such that nonnegative values are obtained.  
After providing background and developing an extremes analogue to weak stationarity, we will construct transformed linear regularly varying analogues to ARMA models and study their properties.  
We will show that fitting a transformed-linear model is done easily, and that our fitted model captures the tail dependence in the windspeed data better than a Gaussian or two classical linear regularly varying models.

\section{Regularly-Varying Time Series and Weak Tail Stationarity}
\label{sec:regVarTS}

\subsection{Regularly-Varying Time Series}
\label{sec:regularvariation}

Before discussing regularly-varying time series, it is useful to briefly review univariate and multivariate regular variation.
One of many formal definitions says $X$ is a regularly-varying random variable, denoted $X \in RV(\alpha)$, if
$
\text{Pr}\left(|X| >x\right) = L(x)x^{-\alpha}
$
for  $x>0$ and a slowly varying function $L$, that is, $\lim_{x \rightarrow \infty}$ $L(ax)/L(x)$ $=1$ for all $a > 0$.
The tail index $\alpha$ describes the power-law-like behavior of the tail.
The upper and lower tails can be described individually as
\begin{eqnarray}
    \label{eq:upperAndLowerTails}
    \text{Pr}\left(X >x\right) = u L(x)x^{-\alpha} \mbox{ and } \text{Pr}\left(X < -x\right) = v L(x)x^{-\alpha}, 
\end{eqnarray}
where $u, v \geq 0, u+v = 1.$

$X$ is a $p$-dimensional multivariate regularly-varying random vector $(X \in RV^p(\alpha))$ if there exists a non-trivial limit measure $\nu_{X}(\cdot)$ and function $b(s) \rightarrow \infty$ as $s \rightarrow \infty$, such that
\begin{eqnarray} \label{eq1}
s \text{Pr}\left( \frac{X}{b(s)} \in \cdot\right) \xrightarrow{v} \nu_{X}(\cdot) \mbox{ as } s \rightarrow \infty,
\end{eqnarray}
where $\xrightarrow{v}$ denotes vague convergence in $M_+(\mathbb{R}^p \setminus \{ 0 \})$, the space of non-negative Radon measures on $\mathbb{R}^p \setminus \{ 0 \}$ \citep[Section 6]{resnick2007}.
The normalizing function is of the form $b(s) = U(s)s^{1/\alpha}$ where $U(s)$ is slowly varying.
The limiting measure $\nu_{X}$ has scaling property $\nu_{X}(aC) = a^{-\alpha}\nu_{X}(C)$ for any $a > 0$ and any set $C\subset \mathbb{R}^p \setminus \{ 0 \}$.
When described in polar coordinates, $\nu_{X}$ decomposes into independent radial and angular components.
Given any norm, define the unit ball $\mathbb{S}_{p-1}= \{x\in \mathbb{R}^p : \|x\|=1\}$. 
Let $C(r,B) = \{x \in \mathbb{R}^p : \|x\| >r, \|x\|^{-1}x \in B\}$ for some $r > 0$, and some Borel set $B \subset \mathbb{S}_{p-1}$.
Then $\nu_{X}\{C(r,B)\} = r^{-\alpha}H_{X}(B)$ where $H_{X}$ is the angular measure taking values on $\mathbb{S}_{p-1}$. 
Equivalently, $\nu(dr \times dw) = \alpha r^{-\alpha - 1} dr dH_{X} (w)$.
The normalizing function $b(s)$ and measures $\nu_{X}$ or $H_{X}$ are not uniquely defined by (\ref{eq1}), as $b(s)$ can be scaled by any positive constant which can be absorbed into the limiting measure.

The recent volume by \citet{kuliksoulier2020} provides a comprehensive treatment of regularly varying time series.
We restrict attention to univariate time series.
Following \citet[Definition 5.1.1]{kuliksoulier2020}, define a regularly-varying time series as a sequence $\{X_t\}$, $t\in \mathbb{Z}$, of real-valued random variables whose finite-dimensional distributions are regularly-varying.
Letting $X_{t,p} = (X_t, X_{t+1}, \ldots, X_{t+p-1})^T$ for any $t$ and $p > 0$, $X_{t,p}$ is multivariate regularly-varying and there exists a normalizing sequence $b(s) \rightarrow \infty$ as $s \rightarrow \infty$ yielding limit measure $\nu_{X_{t,p}}(\cdot)$ and angular measure $H_{X_{t,p}}$ on $\mathbb{S}_{p-1}$.

Having defined finite-dimensional limiting measures, \citet[Chapter 5]{kuliksoulier2020} go on to define and prove existence of the tail measure $\nu$, which is the infinite-dimensional limiting measure of the time series.
\citet{kuliksoulier2020} then characterize the stochastic properties of the time series by focusing on its behavior respective to $t = 0$.
They define the tail process as a random element $\{Y_t\}$, $t\in \mathbb{Z}$, with values in $E_0=\{y\in(\mathbb{R})^{\mathbb{Z}} : |y_0|>1\}$ and distribution $\eta$ defined by $\eta=\nu(\cdot \cap E_0)$.
\citet{kuliksoulier2020} characterize the time series focusing on the spectral tail process $\{\Theta_t=Y_t/|Y_0|\}$, $t\in \mathbb{Z}$, which is independent of $|Y_0|$.

Rather than fully characterizing the tail dependence via the tail measure $\nu$ or the spectral tail process $\Theta_t$, we will characterize a time series only via pairwise tail dependency summary measures.
While such a specification does not contain the full information of $\nu$ or $\Theta_t$, we believe that a time series model which matches these summary measures of dependence is useful for answering many questions and that it can be difficult to infer the full tail dependence structure from data.
Furthermore, our approach nicely ties to traditional linear time series modeling, which often focuses only on second-order properties.
To do so, we require a notion of weak tail stationarity (Section \ref{sec:stationarity}), contrary to strict stationarity discussed in \citet{kuliksoulier2020}.

\subsection{Consistency between finite-dimensional measures}
\label{sec:consistency}

We present consistency results required of regularly-varying time series.
These consistency results show that our tail dependence summary measure is sensible regardless of the dimension of the random vector under consideration.
Our presentation only requires consideration of the finite-dimensional distributions.

To focus attention on the upper tail, consider non-negative time series $\{X_t\}$, $t\in \mathbb Z$.
Then $X_{t,p}$, taking values in $\mathbb{R}^p_+ : = [0, \infty)^p$, is multivariate regularly-varying with tail index $\alpha$, denoted $X_{t,p} \in RV^p_+(\alpha)$. 
$X_{t,p}$ has limiting measure $\nu_{X_{t,p}}$ for sets in $\mathbb{R}^p_+ \setminus \{ 0\}$ and angular measure $H_{X_{t,p}}$ on $\mathbb{S}^+_{p-1} = \{x\in \mathbb{R}^p_+ : \|x\|=1\}$.

Let $X_{t,p}^{(-i)}$ be the $(p-1)$-dimensional vector obtained by excluding the $i\textsuperscript{th}$ component of $X_{t,p}$. Then for $A_1 \in \mathbb{R}^{i-1}_+$, $A_2 \in \mathbb{R}^{p-i}_+$, and $\{0\} \notin A_1 \times A_2$, consistency across dimensions requires $\nu_{X_{t,p}}(A_1 \times [0, \infty] \times A_2) = \nu_{ X_{t,p}^{(-i)}}(A_1 \times A_2),$
or equivalently, $  \int_{(r, w) \in A_1 \times [0, \infty] \times A_2} \alpha r^{-\alpha - 1} \text{d}r \text{d}H_{X_{t,p}}(w) = \int_{(r, v) \in A_1 \times A_2} \alpha r^{-\alpha - 1} \text{d}r \text{d}H_{X_{t,p}^{(-i)}}(v),$ where $w$ $\in \mathbb{S}^+_{p-1}$ and $v \in \mathbb{S}^+_{p-2}$.
In Proposition \ref{prop_1}, the representation of the lower dimensional marginal angular measure is derived. 
For ease and generality, we temporarily drop time series notation, letting $X$ be a $p$-dimensional random vector. Proofs for all propositions are given in the appendix.

\begin{proposition}
\label{prop_1}
Let $X = (X_1, \cdots, X_p)^T \in RV_+^p(\alpha)$ with limiting measure $\nu_{X}$ on $[0,\infty]^p \setminus \{ 0 \}$ and angular measure $H_{X}$ on $\mathbb{S}^+_{p-1}$. Let $X_{(l)} \in RV_+^l(\alpha)$ be the marginal $l$-dimensional random vector, $l<p$, with limiting measure $\nu_{X_{(l)}}$ on $[0,\infty]^l \setminus \{ 0 \}$ and angular measure $H_{X_{(l)}}$ on $\mathbb{S}^+_{l-1}$. Let $A_{(l)}(r,B_{l-1}) = \{ X_{(l)} \in \mathbb{R}_+^l: \| X_{(l)} \| > r, \| X_{(l)} \|^{-1} X_{(l)} \in B_{l-1} \}$ where $B_{l-1} \subset \mathbb{S}^+_{l-1}$. Let $A^*_{(l)}(r,B_{l-1}) = \{ X \in \mathbb{R}_+^p: \| X_{(l)} \| > r, \| X_{(l)} \|^{-1} X_{(l)} \in B_{l-1} \}$. Then,
\begin{eqnarray} \label{prop1}
H_{X_{(l)}}(B_{l-1}) = \int_{w \in \mathbb{S}^+_{p-1}:\| w_{(l)} \|^{-1} w_{(l)} \in B_{l-1}} \|w_{(l)} \|^{\alpha} \text{d}H_{X}(w).
\end{eqnarray}
\end{proposition}

\subsection{Tail Stationarity defined via the Tail Pairwise Dependence Function}
\label{sec:stationarity}

We define a notion of weak tail stationarity for regularly-varying time series, analogous to  weak or second-order stationarity of non-extreme time series discussed in \citet{brockwelldavis1991} and elsewhere.
Assuming a constant mean, classical time series analysis focuses on the ACVF which summarizes pairwise dependence, and weak stationarity implies the ACVF is a function only of lag.

For tail stationarity, we likewise require a measure of pairwise dependence.
Many pairwise extremal dependence measures have been suggested.  For example, $\chi(h)$ is a specific example of the more general extremogram of \citet{davis2009}.
However, unlike the ACVF, $\chi(h)$ is not nonnegative-definite. 
So that our dependence measure has properties similar to covariance, we will henceforth assume $\{X_t\}$ is a regularly varying time series with $\alpha = 2$ and we will use the $L_2$ norm to define the angular measures $H_{ X_{t,p}}$ which are defined on $\mathbb{S}^+_{p-1} = \{x\in \mathbb{R}^p_+ : \|x\|_2=1\}$.
Considering the vector $(X_t, X_{t+h})^T$, 
define the tail pairwise dependence function (TPDF) as
\begin{eqnarray} \label{eq5}
\sigma(X_t, X_{t+h}) = \int_{\mathbb{S}^+_{1}} w_t w_{t+h}\text{d}H_{X_t, X_{t+h}}(w). 
\end{eqnarray}
A non-negative regularly varying time series is weakly tail stationary if $\sigma(X_t, X_{t+h}) = \sigma(h)$ for all $t$. 

The TPDF is essentially the extremal dependence measure defined by \citet{larssonresnick2012}, but for the special case of $\alpha=2$ and using $L_2$-norm to define the angular measure.
As we will show in Section \ref{sec:mainfty}, these restrictions will be important to link the TPDF to an innerproduct.
We also do not impose that the angular measure be a probability measure on $\mathbb{S}^+_1$, as this would invalidate the relationship between angular measures as given in (\ref{prop1}). 
The assumption that $\alpha = 2$ may seem restrictive; however, we will show in Section \ref{sec:application} that our models are widely applicable.

In the finite dimensional setting, \citet{cooleythibaud2019} constructed the tail pairwise dependence matrix (TPDM) defined in terms of the $p$-dimensional joint angular measure of random vector $X \in RV^p_+(2)$, and showed that this matrix was non-negative definite. In the time series setting, there is no ``correct" higher dimension to consider and therefore we consider the bivariate angular measures as in (\ref{eq5}).
Recall that a real-valued function $\kappa$ defined on the integers is non-negative definite if the matrix $K = \{\kappa(i-j)\}_{i,j =1}^p$ is non-negative definite for any $p$; i.e., $a^TKa \ge 0$ for all $a \in \mathbb{R}^p \setminus \{0\}$.
Proposition \ref{prop_2} will first show that the TPDF defined in terms of the bivariate angular measure is equivalent to that defined in terms of the higher dimensional angular measure. 
We again speak of a general $p$-dimensional vector $X$.

\begin{proposition}
\label{prop_2}
Let $X = (X_1, \cdots, X_p)^T \in RV_+^p(2)$, $p>2$ with limiting measure $\nu_{X}$ on $[0,\infty]^p \setminus \{ 0 \}$ and angular measure $H_{X}$ on ${\mathbb{S}}^+_{p-1}$. Let $(X_i, X_j)^T \in RV_+^2(2)$ be a marginal random vector with $\nu_{(X_i, X_j)}$ on $[0,\infty]^2 \setminus \{ 0 \}$ and $H_{(X_i, X_j)}$ on ${\mathbb{S}}^+_{1}$. Then,
\begin{align}
\sigma(X_i, X_j) = \int_{v \in {\mathbb{S}}^+_1} v_iv_j \text{d}H_{(X_i, X_j)}(v) = \int_{w \in {\mathbb{S}}^+_{p-1}} w_iw_j \text{d}H_{X}(w).
\end{align}
\end{proposition}

Using Proposition \ref{prop_2}, we can show the TPDF $\sigma(h)$ of a weakly tail stationary time series $\{X_t\}$ is non-negative definite. Let $p>0$ and $a = (a_1, \ldots, a_p)^T \in \mathbb{R}^p \setminus \{0\}$ be given. Then $\sum_{i,j = 1}^p a_i \sigma(i - j) a_j = a^T \Sigma_{X_{1,p}} a$, where $X_{1,p} = (X_1, \ldots, X_p)^T$, and $\Sigma_{X_{1,p}}$ is the matrix whose $i,j$th element is $\int_{w \in \mathbb{S}^+_{p-1}} w_i w_j \text{d}H_{X_{1,p}}(w)$. Let $m=H_{X}(\mathbb{S}^+_{p-1})$ and let $w$ be a random vector such that $\text{Pr}(w \in B_{p-1}) = m^{-1}H_{X_{1,p}}(B_{p-1})$ for any set $B_{p-1} \in \mathbb{S}^+_{p-1}$. Then by definition, $\Sigma_{X _{1,p}} = m\text{E}(ww^T)$, and thus for any vector $a \in \mathbb{R}^p \setminus \{0\}$, $a^T\Sigma_{X_{1,p}}a = m\text{E}(a^Tww^Ta) \ge 0$. The inequality becomes strict if no element of $w$ is a linear combination of the others.

In the finite dimensional case, \citet[Proposition 5]{cooleythibaud2019} additionally show that the TPDM is a completely positive matrix, that is, there exists a $p\times q_*, q_*<\infty$, non-negative matrix $A_*$ such that $\Sigma_{X_{1,p}} = A_*A_*^T$.
This notion can be extended to the idea of a completely positive function. 
Analogous to non-negative definiteness, we say that a real-valued function $\kappa$ defined on the integers is completely positive if the matrix $K = \{\kappa(i-j)\}_{i,j =1}^p$ is completely positive for any $p$.
Thus, the TPDF $\sigma(\cdot)$ is a completely positive function since $\Sigma_{X_{1,p}}$ is a completetly positive matrix by Proposition 5 of \citet{cooleythibaud2019}.

Proposition \ref{prop_2} also gives a way of defining the TPDF at lag 0.
By (\ref{eq1}),
\begin{eqnarray}
    \label{eq:tpdf0}
\lim \limits_{s\to \infty}  s \text{Pr} \left( \frac{X_1}{b(s)} > c \right) 
  = \int_{\mathbb{S}^+_{p-1}} \int_{c/w_1}^\infty 2r^{-3} \text{d}r \text{d}H_{X_{1,p}}(w)
  = c^{-2} \int_{\mathbb{S}^+_{p-1}} w_1^2 \text{d}H_{X_{1,p}}(w)
  = c^{-2} \sigma(0).
\end{eqnarray}
It is perhaps useful to think of the TPDF similarly to the ACVF in that its value at lag $h$ can most easily be interpreted with knowledge of the relative scale provided by $\sigma(0)$, as $\sigma(h)/\sigma(0) \in [0,1]$.
For a general (perhaps non-stationary) regularly varying time series, we will find it useful to rewrite the left expression in (\ref{eq:tpdf0}) by recalling that  $b(s)=U(s)s^{1/2}$ and letting $x=cU(s)s^{1/2}$ to obtain
\begin{eqnarray}\label{eq:tailRatio}
\lim \limits_{x \to \infty} \frac{\text{Pr}(X_t > x)}{x^{-2}L(x)} = k_t,
\end{eqnarray}
where $L(x)$ is a slowly varying function given by $L(x)=(U(s))^{2} = (b(s))^{2}s^{-1}$.
We refer to the left hand side of (\ref{eq:tailRatio}) as the tail ratio of $X_t$.
Just as scale information can be passed between $b(s)$ and $\nu_{ X}$ in (\ref{eq1}), there is ambiguity in (\ref{eq:tailRatio}) as $k_t$ and $L(x)$ can be scaled by any positive number. 
This scale ambiguity has been handled in various ways.
\cite{larssonresnick2012} (see also \cite{kimkokoszka2020}) impose that the angular measure be a probability measure, a restriction we find inconvenient for consistency across dimensions.
\cite{cooleythibaud2019} assumed a Pareto tail letting $b(s) = s^{1/2}$ pushing all scale information into the angular measure.
In Section 3, we will handle this ambiguity as the normalizing sequence $x^{-2}L(x)$ can be defined by the generating noise sequence.

\section{Transformed-Linear Regularly-Varying Models}
\label{sec:models}

\subsection{Previous Linear Constructions of Regularly-Varying Time Series}
\label{sec:stationarity_classical}
Researchers have been studying linear regularly-varying time series models for decades.
Generally, $\{X_t\}$ is defined as
\begin{eqnarray} \label{eq6}
X_t = \sum_{j=-\infty}^{\infty} \psi_j  Z_{t-j}, 
\end{eqnarray}
where $\{Z_t\}, t \in \mathbb{Z}$ is an iid sequence of regularly-varying random variables with tail index $\alpha$.
If $\exists$ $\delta \in (0,\alpha) \cap [0,1]$ such that $\sum_{j=0}^{\infty} |\psi_j|^\delta < \infty$, then $\{X_t\}$ can be shown to converge with probability 1 \citep[][Section 7.2]{embrechtsetal1997}.
Because $\{Z_t\}$ is iid, $\{X_t\}$ is strictly stationary.

If $\{Z_t\}$ has tail index $0 < \alpha < 2$, characterization of dependence via the ACVF is not possible. 
Authors still frequently summarize the dependence structure of $\{X_t\}$ by pairwise summary metrics.
For a linear time series with infinite variance, \citet[][Section 13.3]{brockwelldavis1991}
define the dependence metric $\rho(h) := (\sum_{j=0}^\infty \psi_j\psi_{j+h})/(\sum_{j=0}^\infty \psi_j^2), h = 1, 2, \cdots$.

Much of the previous work considers regularly-varying time series $\{X_t\}$ which take both positive and negative values.
One could use (\ref{eq6}) to construct nonnegative time series by setting $v=0$ in the tail balance condition and restricting $\psi_j \geq 0$ for all $j$.
We will show in Section \ref{sec:application} that transformed-linear constructions which allow for negative coefficients but which still result in nonnegative regularly varying time series allow for more flexibility than standard linear constructions restricted to have positive coefficients.

\subsection{Transformed-Linear Regularly-Varying Time Series}

Reconciling linear operations with positive random variables or vectors is not simple. For example, regular variation restricted to the positive orthant cannot accommodate multiplication by a negative number. Max-linear approaches (e.g., \citet{davisresnick1989}, \citet{strokorbschlather2015}) have been used to build max-stable models, but linear-algebra operations do not have direct analogues due to the maximum operation.

In the finite dimensional setting, \citet{cooleythibaud2019} propose a way to link traditional linear algebra operations to regular variation on the positive orthant.
For two vectors $X_1$ and $X_2$ $\in \mathbb{R}^p_+$ and $f: \mathbb{R} \mapsto \mathbb{R}_+$ applied componentwise to vectors, transformed-linear summation is $X_1 \oplus X_2 : = f\{f^{-1}(X_1) + f^{-1}(X_2)\}$ and scalar multiplication is $a \circ X_1 : = f\{af^{-1}(X_1)\}$ for $a\in\mathbb{R}$. 
Using the transform $f(y)=\text{log}\{1 + \text{exp}(y)\}$, \citet{cooleythibaud2019} apply transformed-linear operations to $X \in RV^p_+(\alpha)$ and show that regular variation is ``preserved", so long as the lower tail condition
\begin{eqnarray} \label{lowerTail}
s\text{Pr}\left[X_i \le \text{exp}\{-kb(s)\}\right] \to 0, \; k>0, \; i = 1,..., p; \; s \rightarrow \infty,
\end{eqnarray}
is met. If $X_1, X_2 \in RV^p_+(\alpha)$ are independent, both with normalizing function $b(s)$ and respective limiting measures $\nu_{X_1}$ and $\nu_{X_2}$, then
\begin{eqnarray}
    \label{eq:addition}
    s \text{Pr}\left(\frac{ X_1 \oplus  X_2}{b(s)} \in \cdot\right) \stackrel{v}{\rightarrow} \nu_{X_1}(\cdot) + \nu_{X_2}(\cdot), \mbox{ and }  
\end{eqnarray}
\begin{eqnarray}
    s \text{Pr}\left( \frac{a \circ X_1}{b(s)} \in \cdot \right) \stackrel{v}{\rightarrow} 
    \left\{ 
    \begin{array}{c l}
    a^\alpha \nu_{X_1} (\cdot) & \mbox{ if } a > 0,\\
    0 & \mbox{ if } a \leq 0,
    \end{array}
    \right.
    \mbox{ as } s\rightarrow \infty.
    \label{eq:multiplication}
\end{eqnarray}
The important property of the transform $f$ is that it and its inverse have negligible effect on large values: $\lim_{y \rightarrow \infty} f(y)/y = \lim_{x \rightarrow \infty} f^{-1}(x)/x = 1$. 

For application to extremes, we extend familiar linear time series models by considering transformed-linear combinations of regularly-varying terms. We say the time series $\{X_t\}$ is a transformed-linear process of regularly-varying terms if for all $t$, it can be represented as
\begin{eqnarray} \label{eq15b}
X_t = \bigoplus_{j=-\infty}^{\infty}\psi_j \circ Z_{t-j},
\end{eqnarray}
where $\sum^{\infty}_{j=-\infty}|\psi_j|<\infty$, and $\{Z_t\}$ is a noise sequence of independent and tail stationary $RV_+(2)$ random variables. Henceforth, we assume normalizing functions $b(s)$ and $x^{-2} L(x)$ are such that $\lim_{s \rightarrow \infty} s\text{Pr}\left\{Z_t/b(s) > c\right\} = c^{-2}$ and $\lim_{x \rightarrow \infty} P(Z_t > x) / \{x^{-2}L(x)\} = 1$.
We also assume that lower-tail condition (\ref{lowerTail}) holds for $Z_t$'s.
The TPDF of $\{Z_t\}$ is $\sigma_{\{Z_t\}}(0) = 1$ and $\sigma_{\{Z_t\}}(h) = 0$ for all $h \neq 0$.
As $\alpha = 2$, the condition $\sum^{\infty}_{j=-\infty}|\psi_j|<\infty$ guarantees that the infinite transformed-sum in (\ref{eq15b}) converges with probability $1$, which we show in Section \ref{sec:mainfty}.
As defined, $\{X_t\}$ is not strictly stationary because the noise sequence $\{Z_t\}$ is not required to be identically distributed.
In fact so long as $\sigma(h) = 0 \;$ for all $h \neq 0$, $\{Z_t\}$ is not really required to be independent for the following results to hold; however, there seems little practical value in imagining time series constructed from such noise sequences.

\subsection{Transformed Regularly-Varying MA($q$) Process}

$X_t$ is a transformed regularly-varying moving average (MA) process of order $q$ if for $\theta_j \in \mathbb{R}$, $\theta_q>0$,
\begin{eqnarray} \label{eq16}
X_t = Z_t \oplus \theta_1 \circ Z_{t-1} \oplus \theta_2 \circ Z_{t-2} \oplus \cdots  \oplus \theta_q \circ Z_{t-q}.
\end{eqnarray}
To show that the transformed sum in (\ref{eq16}) exists, we need to show that it takes a finite value with probability 1; i.e., $\lim \limits_{x\to\infty} \text{Pr}(X_t >x) = 0.$
We can easily extend (\ref{eq:addition}) and (\ref{eq:multiplication}) to the finite linear combination of $q$ $RV_+(2)$ independent variables to give that $X_t \in RV_+(2)$ and
\begin{eqnarray} \label{eq22}
\lim_{s \rightarrow \infty} s\text{Pr}\left(\frac{X_t}{b(s)} > c \right) 
= \lim_{s \rightarrow \infty} s\text{Pr}\left(\frac{(\bigoplus^q_{j=0}\theta_j \circ Z_{t-j})}{b(s)} > c \right) 
= c^{-2} \sum^q_{j=0}\theta_j^{(0)^{2}},
\end{eqnarray}
where $a^{(0)} = \max(a, 0)$ for $a \in \mathbb{R}$. Existence follows as (\ref{eq22}) implies 
\begin{eqnarray} \label{eq26}
\text{Pr}(X_t > x) \sim x^{-2}L(x)\sum^q_{j=0}\theta_j^{(0)^{2}} \rightarrow 0 \text{ as } x\to\infty.
\end{eqnarray}
Rearranging (\ref{eq26}), the tail ratio is
$
\lim \limits_{x \to \infty} \text{Pr}(X_t > x)/\{x^{-2}L(x)\} = \lim \limits_{x \to \infty} \text{Pr}(X_t > x)/\text{Pr}(Z_1 > x) = \sum^q_{j=0}\theta_j^{(0)^{2}}. \nonumber
$

To get the TPDF, let $h > 0$ and let $X_t$ and $X_{t+h}$ be two elements of the series in (\ref{eq16}), i.e.,
\begin{align}
X_t & = Z_t  \oplus \theta_1 \circ Z_{t-1}  \oplus \theta_2 \circ Z_{t-2} \oplus \cdots  \oplus \theta_q \circ Z_{t-q}, \nonumber \\
X_{t+h} & = Z_{t+h} \oplus \theta_1 \circ Z_{t+h-1} \oplus \theta_2 \circ Z_{t+h-2} \oplus \cdots  \oplus \theta_q \circ Z_{t+h-q}. \nonumber
\end{align}
Let $Z = (Z_{t+h}, Z_{t+h-1}, \cdots, Z_{t-q})^T$ be the $h+q$ dimensional vector of the regularly-varying noise terms. Let $\theta_t = (0, \cdots, 0, 1, \theta_1, \cdots, \theta_q)^T$ and $\theta_{t+h} = (1, \theta_1, \cdots, \theta_q, 0, \cdots, 0)^T$ be the coefficient vectors for $X_t$ and $X_{t+h}$, respectively, such that, $X_t= \theta_t^T \circ Z$, and $X_{t+h}= \theta_{t+h}^T \circ Z$. By Corollary A2 and Proposition 2 of \citet{cooleythibaud2019}, $(X_t, X_{t+h})^T$ has the joint angular measure $H_{(X_t, X_{t+h})}(\cdot)= \sum^{h+q}_{j=0}\|\theta^{(0)}_{.j}\|^2_2 \delta_{\theta^{(0)}_{.j}/\|\theta^{(0)}_{.j}\|} (\cdot ),$
where $\theta_{.j} = (\theta_{tj}, \theta_{(t+h)j})^T = (\theta_j, \theta_{j+h})^T$, $j=0, \cdots,h+q$; $\theta^{(0)}_{.j} = \text{max}(\theta_{.j}, 0)$ applied component-wise and $\delta$ is the Dirac mass function. By (\ref{eq5}), the TPDF is then given by
\begin{eqnarray}\label{eq27}
\sigma(h) = \sum^{h+q}_{j=0} \left(\frac{\theta_{tj}^{(0)}}{\|\theta^{(0)}_{.j}\|_2}\right) \left( \frac{\theta_{(t+h)j}^{(0)}}{\|\theta^{(0)}_{.j}\|_2} \right) \|\theta^{(0)}_{.j}\|^2_2 = \sum^{h+q}_{j=0}\theta_j^{(0)} \theta_{j+h}^{(0)} = \sum^q_{j=0}\theta_j^{(0)} \theta_{j+h}^{(0)},
\end{eqnarray}
where $\theta_0 = 1$ and we drop the $h$ terms in the last summation since $\theta_j=0$ for all $j>q$. As the TPDF depends only on lag, $\{X_t\}$ is stationary. In defining the MA($q$) process in (\ref{eq16}), we require that $\theta_q>0$ because if $\theta_q\le 0$, the TPDF of the MA($q$) will be the same as that of a lower order MA process with the same $\theta_j$ coefficients, $j<q$.

\subsection{Transformed Regularly-Varying MA($\infty$) Process}
\label{sec:mainfty}

Consider the transformed regularly-varying MA($\infty$) time series model
\begin{eqnarray} \label{eq30}
X_t = \bigoplus \limits_{j=0}^{\infty} \psi_j \circ Z_{t-j},
\end{eqnarray}
where $\psi_j\in\mathbb{R}$ and $\sum_{j=0}^{\infty} |\psi_j| < \infty$. In order to draw on previous results for (standard) linear regularly-varying time series, we consider the time series of preimages $\{Y_t\}$ where $X_t = f \left\{\sum_{j=0}^\infty \psi_jf^{-1}(Z_{t-j})\right\} := f(Y_t).$
Let $Y_t^{(q)}$ be the truncated series of the first $q$ terms of $Y_t$. $Y_t^{(q)}$ is $RV(2)$.

Consider the time series which is the difference in the infinite time series in (\ref{eq30}) and the truncated MA($q$) time series in (\ref{eq16}), as
\begin{eqnarray} \label{eq37}
X_t^{(q)'} = \bigoplus \limits_{j=q+1}^{\infty} \psi_j \circ Z_{t-j} = f \left\{\sum_{j=q+1}^\infty \psi_jf^{-1}(Z_{t-j})\right\} := f(Y_t^{(q)'}).
\end{eqnarray}
Lemma A3.26 in \citet{embrechtsetal1997} implies that, as $x \to \infty$,
\begin{eqnarray}
\text{Pr}(Y_t^{(q)'} > x) \sim x^{-2} L(x) \sum_{j=q+1}^{\infty} |\psi_j|^2 
\implies \lim \limits_{q \to \infty} \text{Pr}(Y_t^{(q)'} > x) \sim \lim \limits_{q \to \infty} x^{-2} L(x) \sum_{j=q+1}^{\infty} |\psi_j|^2 = 0. \nonumber
\end{eqnarray}
By Lemma A2 of \citet{cooleythibaud2019}, the limiting measure of 
$X_t^{(q)'}$
also tends to $0$ as $q \to \infty$ and $X_t^{(q)'} \in RV_+(2)$. Rearranging, we get that the tail ratio of $X_t^{(q)'}$ tends to $0$ as $q \to \infty$. We say that the MA($q$) time series converges to the MA($\infty$) time series in tail ratio, as $q \to \infty$.
Convergence in tail ratio is analogous to mean square convergence in the classical non-heavy-tail case. 
Consequently, the infinite series in (\ref{eq30}) converges since, $\text{Pr}(X_t > x) \sim x^{-2} L(x) \sum_{j=0}^{\infty} \psi_j^{(0)^2} \to 0 \text{, as } x \to \infty.$
Also, taking limit as $q\to\infty$ in (\ref{eq27}), the TPDF is
\begin{eqnarray} \label{eq37e}
\sigma(h) = \sum^{\infty}_{j=0} \psi^{(0)}_j \psi^{(0)}_{j+h} < \infty,
\end{eqnarray}
where $\psi^{(0)}_j =$ max($\psi_j,0$) and $\psi_0 = 1$. Thus the infinite series in (\ref{eq30}) is stationary.

\subsection{Inner Product Space $\mathbb{V}$}

Before we discuss AR and ARMA time series models, we broaden the discussion to better understand why we consider $\alpha=2$ and use the $L_2$ norm.
Consider the space $\mathbb{V} = \{X_t : X_t = \bigoplus^{\infty}_{j=0}\psi_{t,j} \circ Z_j, \sum^{\infty}_{j=0} |\psi_j|<\infty\}$
where $Z_j$'s are independent and tail stationary $RV_+(2)$ random variables with $\text{Pr}(Z_j > x)/(x^{-2}L(x)) = 1$ and $\psi_{t,j}\in \mathbb{R}$.
It can be easily shown that $\mathbb{V}$ is a vector space (Supplementary Material).
For any $X_t \in \mathbb{V}$ we can define a mapping $T:\mathbb{V} \to \ell^1 = \big\{ \{a_j\}^{\infty}_{j=0}, a_j \in \mathbb{R} : \sum^{\infty}_{j=0} |a_j|< \infty \big\}$ such that $T(X_t) = \{\psi_{t,j}\}_{j=0}^{\infty} \in \ell^1$. As $T$ is a linear map and an isomorphism (Supplementary Material), $\mathbb{V}$ is isomorphic to $\ell^1$.

Let $X_t$ and $X_s$ be two elements of vector space $\mathbb{V}$.
We define the inner product between $X_t$ and $X_{s}$ as
\begin{eqnarray} \label{eq:innerprod}
\langle X_t, X_s \rangle := \sum^{\infty}_{j=0} \psi_{t,j}\psi_{s,j}.
\end{eqnarray}
We show (\ref{eq:innerprod}) satisfies the properties of an inner product (Supplementary Material). 
For $X_t \in \mathbb{V}$, the norm 
of $X_t$ is defined as $\left\Vert X_t\right\Vert = \sqrt{\langle X_t, X_t \rangle} = \sqrt{\sum^{\infty}_{j=0} \psi_{t,j}^2},$
which is finite as $\sum^{\infty}_{j=0} |\psi_{t,j}|<\infty$.
\noindent{$X_t, X_s \in \mathbb{V}$} are said to be orthogonal if $\langle X_t, X_s \rangle = \sum^{\infty}_{j=0} \psi_{t,j}\psi_{s,j} = 0.$

If $\{X_t\}$ is an MA($\infty$) time series, $X_t \in \mathbb{V}$ for all $t$.
As $\{X_t\}$ is stationary, it is natural to think of the inner product as a function of lag:
$\gamma(h) = \langle X_t, X_{t + h} \rangle = \sum^{\infty}_{j=0}\psi_j \psi_{j+h}.$
Because we assume $\alpha=2$ and use the $L_2$ norm, the TPDF $\sigma(h)$ is closely related to $\gamma(h)$.
$\gamma(h)$ is equivalent to $\sigma(h)$ if $\psi_j\ge0$ for all $j$.
In a forthcoming paper, we discuss prediction for stationary time series. 
Although $\mathbb{V}$ itself is not a Hilbert space since $\ell^1$ is not complete in the metric induced by the $l^2$ inner product, the set of predictors based on previous $n$ observations is isomorphic to a closed linear subspace of $l^2$ and we can employ the projection theorem.

\subsection{Transformed Regularly-Varying Auto-Regressive Processes}
\label{sec:ar}

Consider the stationary transformed regularly-varying autoregressive (AR) model of order 1 where {$X_t$} is defined as the stationary solution to the equation
\begin{eqnarray} \label{eq38}
  X_t = \phi \circ X_{t-1} \oplus Z_t,      
\end{eqnarray}
where $|\phi| < 1$. Iterating, we obtain $X_t = \phi \circ X_{t-1} \oplus Z_t = \phi \circ (\phi \circ X_{t-2} \oplus Z_{t-1}) \oplus Z_t = \phi^2 \circ X_{t-2} \oplus \phi \circ Z_{t-1} \oplus Z_t = \cdots = \phi^k \circ X_{t-k} \bigoplus_{j=0}^{k-1} \phi^j \circ Z_{t-j}.$
If $|\phi|<1$, the first term becomes small as $k$ increases.

To show $X_t = \bigoplus_{j=0}^\infty \phi^j \circ Z_{t-j}$ is a solution, rewrite in terms of transform $t$ to get
\begin{eqnarray} \label{eq41}
  Z_t = X_t \oplus (-\phi) \circ X_{t-1} 
  = f \left\{f^{-1}(X_t) - \phi f^{-1}(X_{t-1}) \right\}.
\end{eqnarray}
\citet{brockwelldavis1991} define the backward shift operator as $BX_t = X_{t-1}$.
For our transformed-linear operations, we use the same notation to define the transformed backward shift operator as, $B f^{-1}(X_t) = f^{-1}(X_{t-1})$, or equivalently, $f\{B f^{-1}(X_t)\} = X_{t-1}.$
Powers of the operator $B$ are defined as $B^j f^{-1}(X_t) = f^{-1}(X_{t-j})$. Transformed-polynomials in $B$ are isomorphic to polynomial functions of real variables and can be manipulated the same way.
Rewriting (\ref{eq41}) in terms of the operator $B$, we get, $Z_t = f \left\{(1-\phi B)f^{-1}(X_t)\right\} = (1-\phi B) \circ X_t.$
Defining $\phi(B)=1-\phi B$ we get, $Z_t = \phi(B) \circ X_t.$
We want to find $\pi(B)$ such that,
\begin{eqnarray} \label{eq46}
 \pi(B) \circ Z_t = \{\pi(B) \phi(B)\} \circ X_t \text{, and } \pi(B) \text{ and } \phi(B) \text{ are inverses}.
\end{eqnarray}
To find the inverse $\pi(B)$, consider $\phi(z) = 1-\phi z$. It can be shown through a Taylor series expansion that $\pi(z) = \sum_{j=0}^\infty \{z^j \pi^{(j)}(0)\}/(j!) = \sum_{j=0}^\infty \phi^j z^j,$
the series being convergent only if $|\phi z|<1$. In terms of the transformed backward shift operator $B$, $\pi(B) =  \sum_{j=0}^\infty \phi^j B^j$.
Substituting $\pi(B) =  \sum_{j=0}^\infty \phi^j B^j$ in (\ref{eq46}) and solving (refer Appendix \ref{a2}) we get,
\begin{eqnarray} \label{eq52}
 X_t = \bigoplus_{j=0}^\infty \phi^j \circ Z_{t-j}.
\end{eqnarray}
To show uniqueness, given any stationary solution $Y_t$ of (\ref{eq38}), $Y_t = \phi \circ Y_{t-1} \oplus Z_t = \cdots \text{iterating}\cdots = \phi^{k+1} \circ Y_{t-(k+1)} \oplus \bigoplus_{j=0}^{k} \phi^j \circ Z_{t-j}.$
Rearranging, we get,
\begin{eqnarray} \label{eq54}
 Y_t \oplus \bigoplus_{j=0}^{k} (-\phi^j) \circ Z_{t-j} = \phi^{k+1} \circ Y_{t-(k+1)}.
\end{eqnarray}
The right hand side of (\ref{eq54}) is a regularly-varying random variable. The stationary time series $\{Y_t\}$ has a finite tail ratio. The tail ratio of $\phi^{k+1} \circ Y_{t-(k+1)} = (\phi^{2k+2}) \times \text{tail ratio of }Y_{t}$ which is finite. Taking limit as $k \to \infty$,
\begin{eqnarray} \label{eq56}
\lim \limits_{k \to \infty} \left[\text{ tail ratio of }\{Y_t \oplus \bigoplus_{j=0}^{k} (-\phi^j) \circ Z_{t-j}\} \right] = \lim \limits_{k \to \infty} \left[ \phi^{2k+2} \times \text{ tail ratio of }\{Y_t\}\right] = 0,
\end{eqnarray}
as $|\phi|<1$. The limit in (\ref{eq56}) goes to $0$, indicating that $Y_t$ is equal to the tail ratio limit $\bigoplus_{j=0}^\infty \phi^j \circ Z_{t-j}$ and that the process defined by (\ref{eq52}) is the unique stationary solution of (\ref{eq38}).

The TPDF for the AR(1) is then given by, $\sigma_{X_t,X_{t+h}} = \sigma(h) = \sum^{\infty}_{j=0} \phi^{j^{(0)}} \phi^{j+h^{(0)}},$ and the tail ratio for the AR(1) is, $\sigma_{X_t,X_t} = \sigma(0) = \sum^{\infty}_{j=0} (\phi^{j^{(0)}})^2.$

In the case $|\phi|>1$, the series in (\ref{eq52}) does not converge as $\sum_{j=0}^{\infty}|\phi^j|$ does not converge. However, we can write (\ref{eq38}) for $X_{t+1}$ as,
\begin{align}
\phi \circ X_t & = X_{t+1} \oplus (-Z_{t+1})   \nonumber \\
\Rightarrow X_t & = \phi^{-1} \circ \left\{ X_{t+1} \oplus (-Z_{t+1})\right\} = \phi^{-1} \circ X_{t+1} \oplus (-\phi^{-1}) \circ Z_{t+1} \nonumber\\
    & = \cdots = \phi^{-k-1} \circ X_{t+k+1} \oplus (-\phi^{-k-1}) \circ Z_{t+k+1} \oplus \cdots \oplus (-\phi^{-1}) \circ Z_{t+1}, \nonumber
\end{align}
which shows, by the same arguments as earlier that, $X_t =  \bigoplus^{\infty}_{j=1} (-\phi^{-j}) \circ Z_{t+j},$
is the unique stationary solution of (\ref{eq38}) as $\sum^{\infty}_{j=1} |\phi|^{-j} <\infty$. Similar to non-extreme time series notion of causality, we say that $\{X_t\}$ is causal if $X_t$ can be expressed in terms of the current and past values, $Z_s$, $s \le t$. Thus for $\{X_t\}$ defined as a solution to (\ref{eq38}), $\{X_t\}$ is causal if $|\phi| < 1$, $\{X_t\}$ is non-causal if $|\phi| > 1$ and there is no stationary solution if $|\phi| = 1$.

\subsection{Transformed Regularly-Varying ARMA(1,1) Process}
\label{sec:arma11}

Consider the transformed regularly-varying ARMA(1,1) model where {$X_t$} is defined as the stationary solution to the equation
\begin{eqnarray} \label{eq66}
  X_t \oplus (-\phi) \circ X_{t-1}  = Z_t \oplus \theta \circ Z_{t-1},      
\end{eqnarray}
where $\theta + \phi \ne 0$. Using the transformed backward shift operator B as defined in Section \ref{sec:ar}, (\ref{eq66}) can be rewritten as,
\begin{eqnarray} \label{eq67}
  \phi(B) \circ X_t = \theta(B) \circ Z_t,
\end{eqnarray}
where $\phi(B) = 1 - \phi B$ and $\theta(B) = 1 + \theta B$.
Following \citet{brockwelldavis2002}, we investigate the range of values of $\phi$ and $\theta$ for which a stationary solution of the ARMA(1,1) exists. If $|\phi|<1$, we can rewrite (\ref{eq67}) as,
\begin{align} \label{eq68}
    X_t = \left[\{\phi(B)^{-1}\}\theta(B)\right] \circ Z_t = \left\{\pi(B)\theta(B)\right\} \circ Z_t = \psi(B) \circ Z_t,
\end{align}
where $\pi(B) = \sum^{\infty}_{j=0}\phi^jB^j$.
Expanding $\psi(B)$ and substituting in (\ref{eq68}) (refer Appendix \ref{a3}),
\begin{align} \label{eq69}
    X_t & = Z_t \oplus (\phi+\theta) \circ \left(\bigoplus^{\infty}_{j=1}\phi^{j-1} \circ Z_{t-j}\right).
\end{align}
We conclude that the MA($\infty$) process (\ref{eq69}) is the unique stationary solution of (\ref{eq66}), and is causal.

The TPDF for the ARMA(1,1) is complicated, and is developed from (\ref{eq69}) in the Appendix \ref{a5}. In contrast to the MA and AR(1) models where negative parameters $\theta_j$ or $\phi$ do not affect the TPDF because of the zero operation, negative parameter values do influence the TPDF of the transformed-linear ARMA(1,1) which does not match the ACVF of a non-extreme ARMA(1,1).

If $|\phi|>1$, we can express $1/\phi(z)$ as a power series of $z$ with absolutely summable coefficients by expanding in powers of $z^{-1}$, giving $1/\phi(z) = -\sum^{\infty}_{j=1}\phi^{-j}z^{-j}.$
Applying the same argument as in the case where $|\phi|<1$, we can obtain the unique stationary solution of (\ref{eq66}).
Letting $\zeta(B) = -\sum^{\infty}_{j=1}\phi^{-j}B^{-j}$, and applying $\zeta(B)$ to both sides of (\ref{eq67}) (refer Appendix \ref{a4}), we get,
\begin{align} \label{eq71}
    X_t & = (-\phi^{-1}\theta) \circ Z_t \oplus -(\phi+\theta) \circ \left(\bigoplus^{\infty}_{j=1}\phi^{-j-1} \circ Z_{t+j}\right).
\end{align}
In this case the solution is noncausal. If $\phi = \pm1$, there is no stationary solution of (\ref{eq66}).

We can show that the ARMA(1,1) process in (\ref{eq66}) is invertible, i.e., $Z_t$ is expressible in terms of current and past values, $X_s$, $s \le t$. Let $\xi(z) = 1/\theta(z) = \sum^{\infty}_{j=0}(-\theta)^jz^j$ with absolutely summable coefficients. Applying $\xi(B)$ to both sides of (\ref{eq67}), $Z_t = \{\xi(B)\phi(B)\} \circ X_t = \pi(B) \circ X_t$,
where $\pi(B) = \sum^{\infty}_{j=0}\pi_jB^j = (1 - \theta B + (-\theta)^2B^2 + \cdots ) (1-\phi B).$ As earlier,
  $Z_t = X_t \oplus -(\phi+\theta) \circ \{\bigoplus^{\infty}_{j=1}(-\theta)^{j-1} \circ X_{t-j}\}.$ 
Following an argument like the one used to show noncausality when $|\phi|>1$, the ARMA(1,1) process is noninvertible when $|\theta| > 1$, since then $Z_t$ is expressed in terms of current and future values, $X_s$, $s \ge t$, by
  $Z_t = (-\phi \theta^{-1}) \circ X_t \oplus (\phi+\theta) \circ \{\bigoplus^{\infty}_{j=1}(-\theta)^{-j-1} \circ X_{t+j}\}.$ 
Analogous to the non-extreme time series case, if $\theta = \pm1$, the ARMA(1,1) process is invertible in the more general sense that $Z_t$ is a tail ratio limit of finite transformed-linear combinations of $X_s$, $s \le t$, although it cannot be expressed explicitly as an infinite transformed-linear combination of $X_s$, $s \le t$. A noncausal or noninvertible ARMA(1,1) process $\{X_t\}$ can be re-expressed as a causal and invertible ARMA(1,1) process relative to a new regularly-varying noise sequence $\{Z^*_t\}$. Thus, we can restrict attention to causal and invertible ARMA(1,1) models with $|\phi|<1$ and $|\theta|<1$.
This is also valid for higher-order ARMA models.

\subsection{Transformed Regularly-Varying ARMA($p,q$) Process} \label{sec:armapq}

Consider the stationary transformed regularly-varying ARMA($p,q$) model where {$X_t$} is defined as the stationary solution to the equation
\begin{eqnarray} \label{eq74}
  X_t \oplus (-\phi_1) \circ X_{t-1} \oplus \cdots  \oplus (-\phi_p) \circ X_{t-p}  = Z_t \oplus \theta_1 \circ Z_{t-1} \oplus \cdots  \oplus \theta_q \circ Z_{t-q},
\end{eqnarray}
where polynomials ($1-\phi_1z-\cdots -\phi_pz^p$) and ($1+\theta_1z+\cdots +\theta_qz^q$) have no common factors.
We can express (\ref{eq74}) in terms of the transformed backward shift operator as
\begin{eqnarray} \label{eq74a}
  \phi(B) \circ X_t =  \theta(B) \circ Z_t,
\end{eqnarray}
where $\phi(z)=1-\phi_1z-\cdots -\phi_pz$ and $\theta(z)=1+\theta_1z-\cdots -\theta_qz$.
$\{X_t\}$ is said to be a transformed regularly-varying autoregressive process of order $p$ (or AR($p$)) if $\theta(z) = 1$, and a transformed regularly-varying moving average process of order $q$ (or MA($q$)) if $\phi(z) = 1$.

In Section \ref{sec:arma11} we showed that a unique stationary solution exists for the ARMA($1,1$) if and only if $\phi_1\ne\pm 1$. Equivalently, the AR polynomial $\phi(z)=1-\phi(z)\ne 0$ for $z=\pm 1$. The analogous condition for the general ARMA($p,q$) process is $\phi(z)=1-\phi_1z-\cdots -\phi_pz \ne 0$ for all complex $z$ with $|z| = 1$. 
As discussed in \citet[Section 3.1]{brockwelldavis2002}, $z$ could be complex, since the zeros of a polynomial of degree $p > 1$ may be either real or complex.

If $\phi(z)\ne0$ for $z = \pm 1$, then there exists $\delta>0$ such that $\zeta(z)=\phi(z)^{-1}=\sum^{\infty}_{j=-\infty}\zeta_jz^j$ for $1-\delta < |z| < 1+\delta$, and $\sum^{\infty}_{j=-\infty}|\zeta_j|<\infty$. We can then define $1/\phi(B)$ as the linear filter with absolutely summable coefficients, i.e., $1/\phi(B) = \sum^{\infty}_{j=-\infty}\zeta_jB^j.$
Applying $\zeta(B)$ to both sides of (\ref{eq74a}), we get,
\begin{eqnarray}
 X_t = \{\zeta(B)\phi(B)\} \circ X_t = \{\zeta(B)\theta(B)\} \circ Z_t = \psi(B) \circ Z_t = \bigoplus^{\infty}_{j=-\infty}\psi_j\circ Z_{t-j},
\end{eqnarray}
where $\psi(z)=\zeta(z)\theta(z)=\sum^{\infty}_{-\infty}\psi_jz^j$. By similar argument as in Section \ref{sec:arma11}, we can show that $\psi(B)\circ Z_t$ is the unique stationary solution of (\ref{eq74}).

In Section \ref{sec:arma11} we saw that the ARMA(1,1) process is causal if and only if $|\phi_1|<1$. Equivalently, the ARMA($p,q$) process is causal, i.e., $X_t$ can be represented as $\bigoplus^{\infty}_{j=0}\psi_jZ_{t-j}$, with $\sum^{\infty}_{j=0}|\psi_j|<\infty$, if $\phi(z)\ne 0$ for $|z|\le 1$, i.e., the roots of the AR polynomial should lie outside the unit circle. Similarly, we saw that the ARMA(1,1) process is invertible, i.e., $Z_t$ can be represented as $\bigoplus^{\infty}_{j=0}\pi_jX_{t-j}$, with $\sum^{\infty}_{j=0}|\pi_j|<\infty$, if and only if $|\theta_1|<1$. Equivalently, the ARMA($p,q$) process is invertible if $\theta(z)\ne 0$ for $|z|\le 1$, i.e., the roots of the MA polynomial should lie outside the unit circle.

As shown in \citet[Section 3.1]{brockwelldavis2002}, $\{\psi_j\}$ can be determined by, $\psi_j - \sum^p_{k=1}\phi_k\psi_{j-k} = \theta_j, \quad j = 0,1,...,$
where $\theta_0=1$, $\theta_j=0$ for all $j>q$ and $\psi_j=0$ for all $j<0$.
Similarly, $\{\pi_j\}$ can be determined by, $\pi_j + \sum^q_{k=1}\theta_k\pi_{j-k} = -\phi_j, \quad j = 0,1,...,$
where $\phi_0=-1$, $\phi_j=0$ for all $j>p$ and $\pi_j=0$ for all $j<0$.
By the causal representation of the ARMA($p,q$), its TPDF is given by (\ref{eq37e}).

\section{Application to Santa Ana Winds}
\label{sec:application}

\subsection{Data and Preprocessing}
\label{sec:santaana}

Most regularly varying time series models have been constructed with the aim of modelling heavy-tailed data.
If $\{X_t^{(orig)}\}$ is a nonnegative regularly varying time series with tail index $\alpha$, letting $X_t = (X_t^{(orig)})^{\alpha/2}$ yields a time series with tail index 2, and a transformed linear model could be suitable for capturing tail dependence after this simple power transformation.
Because regular variation provides a dependence model which focuses on tail behavior and allows for asymptotic dependence, we think our models can be widely employed to capture dependence in the upper tail.  
As mentioned in Section \ref{sec:introduction}, the windspeed data appears to exhibit very strong tail dependence at short lags, and we feel that an asymptotically dependent model is well-suited to capture this dependence.  
However, the windspeed data are not heavy-tailed.
To model this windspeed data, we will use our transformed-linear models to capture the tail dependence of this data after it has been marginally transformed to have regularly-varying tails with $\alpha = 2$.  
Separating the marginal distribution from the dependence structure is justified by Sklar's theorem (\citet{sklar1959}, see also \citet[Proposition 5.15]{resnick1987}), and such marginal transformations are relatively common in extremes (refer \citet{smithetal} for an example in the time series case).

We return to the March AFB hourly windspeed data for the years 1973 - 2019, first introduced in Section \ref{sec:introduction}.
We choose to focus only on the autumn season (September 22 - December 22) as this is the period when fire risk due to the Santa Ana winds phenomenon is greatest. Our data set consists of 103,630 hourly observations.

Let $\{ x_t^{(\text{orig})}\}$ be the hourly windspeed anomalies after removal of the diurnal cycle. Based on mean residual life plots \citep[Section 4.3]{coles2001}, we fit a generalized Pareto distribution (GPD) to the upper $2.5\%$ of $x_t^{(\text{orig})}$.
The windspeed anomalies appear to have a bounded tail (shape parameter estimate: $\hat \xi$ = -0.1 (se = 0.02)).
Letting $\hat{\mu}$, $\hat{\psi}$ and $\hat{\xi}$ denote the empirical $.975$ quantile, GPD scale and shape estimates, respectively, our estimated marginal distribution is
\begin{align}
\hat{F}(x) & = 
                \begin{cases}
                (n+1)^{-1} \sum^n_{t=1} \mathbb{I}(x_t^{(\text{orig})} \le x) & \text{for } x \le \hat{\mu},\\
                1 - 0.025\{1 + \hat{\xi}(x -\hat{\mu})/\hat{\psi}\}^{-1/\hat{\xi}} & \text{for } x > \hat{\mu}.
                \end{cases} \nonumber
\end{align}
To obtain data which fits into our regular-variation modeling framework, we define $x_t = G^{-1}\{\hat{F}(x^{(\text{orig})}_t)\}$, where $G(x) = \text{exp}(-x^{-2})$. Thus, our transformed time series $\{x_t\}$ will have a marginal distribution which is approximately Fr\'echet with $\alpha = 2$ and $\sigma(0) = 1$.

\subsection{Determination of Model TPDF's}
\label{sec:modelTPDF}

Our proposed model fitting method is to find the parameters which minimize the squared difference between the empirical and model TPDF's. Due to preprocessing, $\sigma(0)$ is known to be 1.  For lag $h > 0$, the model TPDFs with tail ratio 1 for the three models that we fit to the application data are as follows:
\begin{align}
\text{AR(1): } & \sigma(h)=\text{max}(0, \phi^h), \nonumber \\
\text{MA(1): } & \sigma(h)= 
                \begin{cases}
                \frac{\theta}{1+\theta^2} & \text{if } h = 1, \theta>0\\
                0 & \text{otherwise},
                \end{cases} \nonumber
\end{align}
\begin{align}
\text{ARMA(1, 1): } & \sigma(h)= 
                \begin{cases}
                \frac{(\phi+\theta) \phi^h (1+\phi\theta)}{1+2\phi\theta + \theta^2} & \text{if } \phi > 0, \phi+\theta>0 \\
                0 & \text{if } \phi > 0, \phi+\theta<0 \\
                \frac{(\phi+\theta)^2\phi^h}{1-\phi^4+(\phi+\theta)^2} & \text{if } \phi < 0, \phi+\theta>0, h \text{ is even} \\
                \frac{(\phi+\theta)\phi^{h-1}(1-\phi^4)}{1-\phi^4+(\phi+\theta)^2} & \text{if } \phi < 0, \phi+\theta>0, h \text{ is odd} \\
                \frac{(\phi+\theta)\phi^{h-1}(1+\theta\phi^3)}{1+\phi^2\theta^2+2\phi^3\theta} & \text{if } \phi < 0, \phi+\theta<0, h \text{ is even} \\
                0 & \text{if } \phi < 0, \phi+\theta<0, h \text{ is odd}. \nonumber
                \end{cases} \nonumber
\end{align}

\subsection{Estimation of TPDF, Model Fitting, and Comparison}
\label{sec:estimation}

To estimate the TPDF, we use the estimator defined in \citet{cooleythibaud2019} in which the true angular measure is replaced by an empirical estimate. 
Assuming $(x_t,x_{t+h})^T, (t = 1, \cdots, n-h)$ be lag-$h$ pairs of observations from a tail stationary time series $\{X_t\}$ and letting $r_t = \|(x_t,x_{t+h})^T\|_2$, and $w_t = (w_t, w_{t+h})^T = (x_t, x_{t+h})^T/r_t$, 
the TPDF estimator is defined as
\begin{eqnarray} \label{eq78a}
  \hat{\sigma}(h) 
  = 2 \int_{\Theta^+_{1}} w_tw_{t+h}d\hat{N}_{X_t,X_{t+h}}(w) 
  = \frac{2}{\sum^{n}_{t=1}\mathbb{I}(r_t>r_0)} \sum^{n}_{t=1} w_tw_{t+h} \mathbb{I}(r_t>r_0),
\end{eqnarray}
where $r_0$ is some high threshold for the radial components, $H_{X_t,X_{t+h}}(\Theta^+_{1}) = 2$ because $\sigma(0) = 1$, and $N_{X_t,X_{t+h}}(\cdot) = 2^{-1}H_{X_t,X_{t+h}}(\cdot).$

\begin{figure}
\centerline{\includegraphics[width=0.5\textwidth, trim = 0cm 1.5cm 0cm 1.5cm, clip]{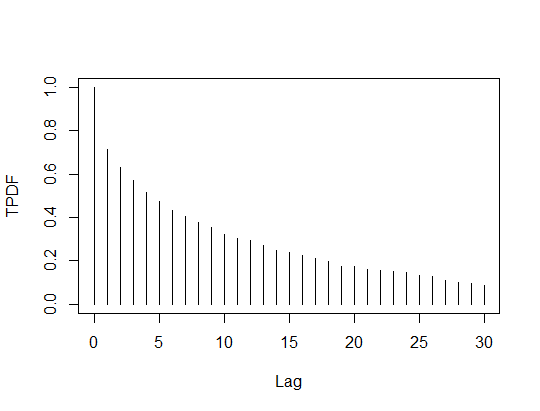}
      \includegraphics[width=0.5\textwidth, trim = 0cm 1.5cm 0cm 1.5cm, clip]{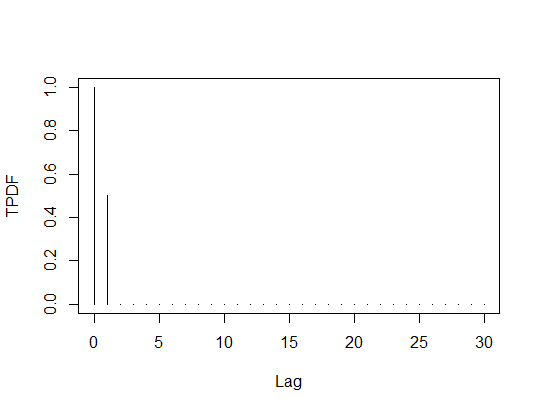}}
      
\centerline{\includegraphics[width=0.5\textwidth, trim = 0cm 0.5cm 0cm 1.5cm, clip]{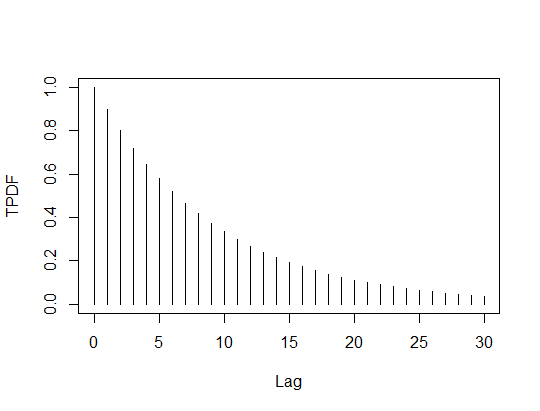}
      \includegraphics[width=0.5\textwidth, trim = 0cm 0.5cm 0cm 1.5cm, clip]{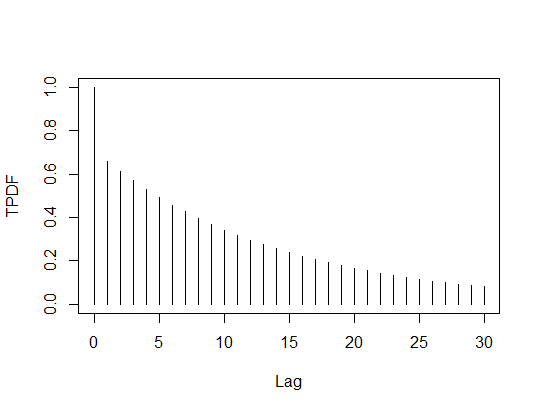}}
    
    \caption{Comparison of estimated TPDF from data (upper left panel) and theoretical TPDF of fitted models: MA(1) (upper right panel), AR(1) (lower left panel) and ARMA(1,1) (lower right panel).}
  \label{Figure2}
\end{figure}

It is known that tail dependence estimates tend to have positive bias in the case of weak dependence \citep{huseretal2016}. Simulation study (refer Supplementary Material) shows that subtracting off the mean of the time series considerably reduces bias in TPDF estimation. We subtract off the mean of the transformed time series $\{x_t\}$ and replace the negative observations by 0. We estimate the TPDF for the first 30 lags using (\ref{eq78a}). The upper left panel of Figure \ref{Figure2} gives the empirical TPDF for the data after bias correction.

Parameter estimates for the three transformed regularly-varying time series models are obtained by numerical least squares optimization. The parameter estimates and sum of squared differences (SS) between the empirical and model TPDF's are given in Table \ref{table_fit}, and the ARMA(1,1) has the lowest SS.
We can also see from Figure \ref{Figure2} that compared to the other two models, the theoretical TPDF of the ARMA(1,1) model (lower right panel) seems to be a closer fit to the empirical TPDF (upper left panel).
\begin{table}[h]
\fontsize{10}{10}\selectfont
 \begin{center}
    \caption{Fitted models}
    \label{table_fit}
        \begin{tabular}{llr}
        \hline
        Model & Parameter estimates & SS   \\ \hline
        MA(1) & $\hat{\theta} = 1$ & 2.71 \\ 
        AR(1) & $\hat{\phi} = 0.9$ & 0.19 \\
        ARMA(1,1) & $\hat{\phi} = 0.93$, $\hat{\theta} = -0.51$ & 0.01 \\ \hline
        \end{tabular}
 \end{center}
\end{table}

\begin{figure}[htb]
\centering
    \includegraphics[width=0.8\textwidth, trim = 0cm 1.5cm 0cm 1cm, clip]{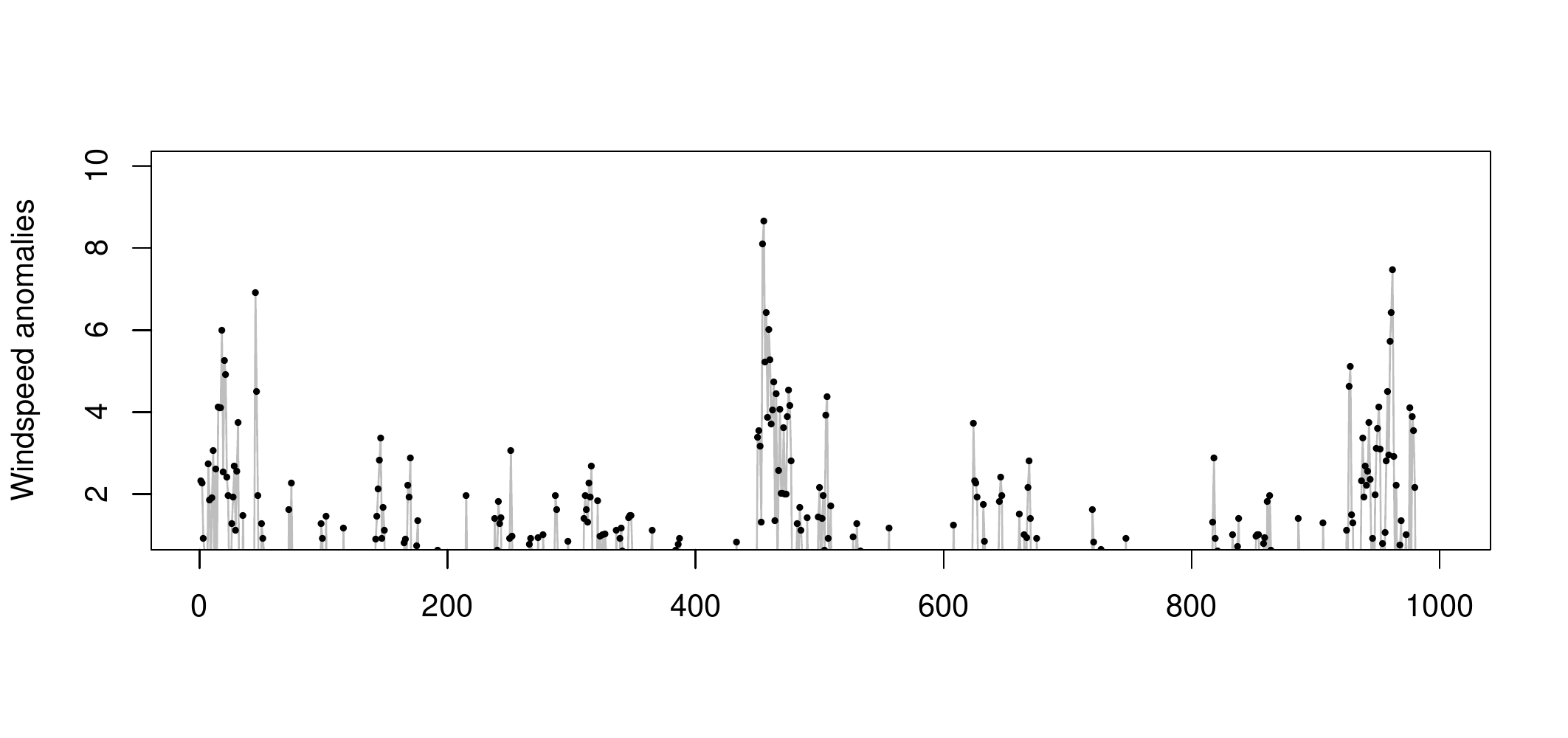}

    \includegraphics[width=0.8\textwidth, trim = 0cm 0.5cm 0cm 1.5cm, clip]{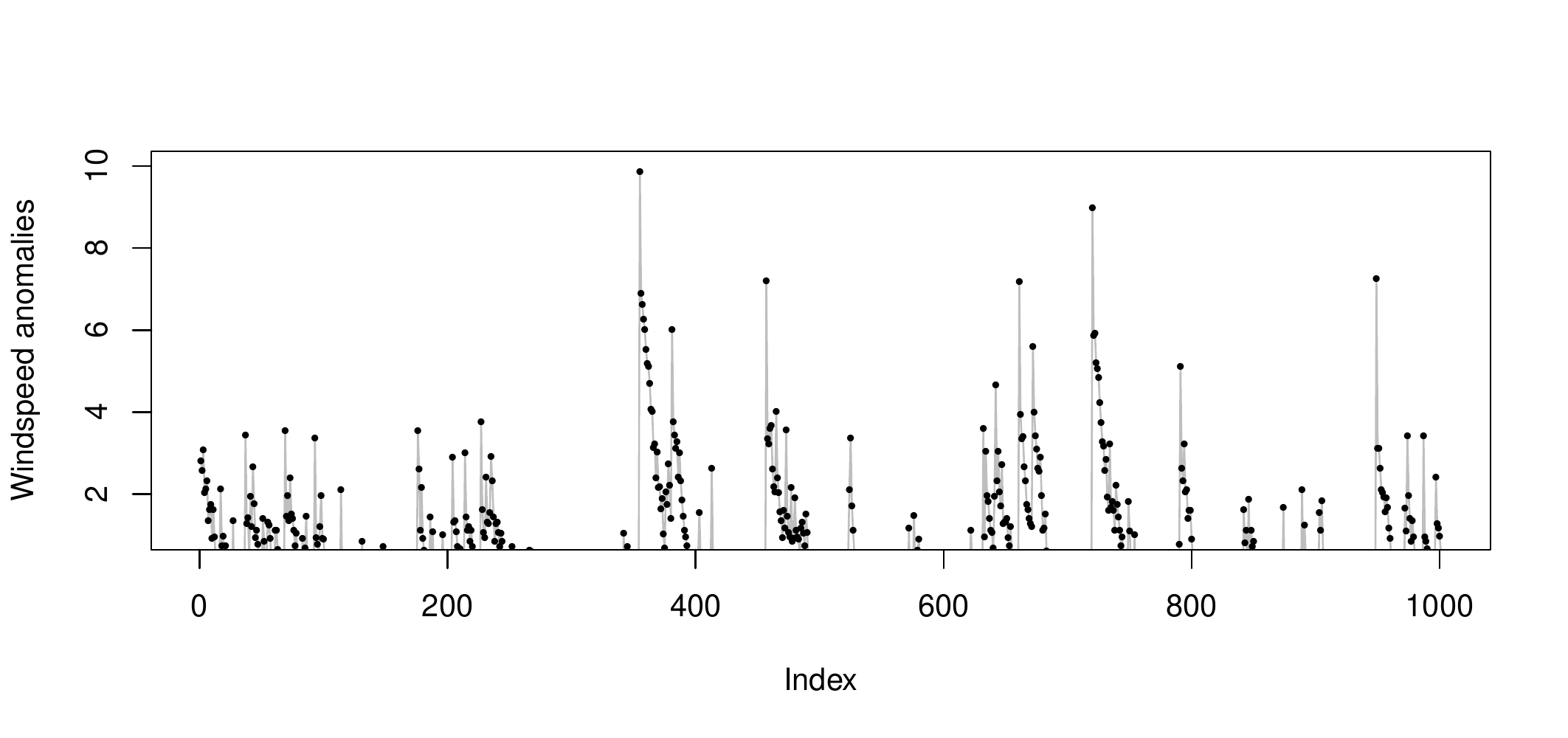}
  \caption{Comparison, above a threshold of 1 m/s, of actual windspeed anomalies time series (upper panel) and realization of synthetic time series generated from fitted transformed regularly-varying ARMA(1,1) model (lower panel), transformed to marginal of original time series.}
  \label{Figure3}
\end{figure}

We generate a realization of the fitted transformed regularly-varying ARMA(1,1) time series model and transform it to the marginal of the observed hourly windspeed anomalies time series. Figure \ref{Figure3} compares the actual time series (upper panel) with the generated synthetic time series (lower panel) above a threshold of 1m/s. The two time series look quite similar to each other above this threshold.

For comparison, we fit three alternative ARMA(1,1) models, each with a two-step procedure.
The first is a linear Gaussian time series model. 
We transform the marginal to be standard normal, estimate the ACVF, and then estimate the ARMA(1,1) parameters. 
The second is a linear regularly-varying time series model which follows the spirit of the models reviewed in Section \ref{sec:stationarity_classical} that take values in $\mathbb{R}$. 
To be comparable, we perform a marginal transformation so that the data is unit Fr\'echet with $\alpha=2$ in both directions. 
We then estimate the TPDF for this transformed data, which differs because the bivariate angular measure is not restricted to the positive orthant. 
For this model, the TPDF's summarized tail dependence is symmetric because it assesses both the upper and lower tail. 
Dependence in the lower tail of the windspeed anomalies is not as strong as in the upper tail ($\hat \chi(1) \approx 0.4$ for the negated anomaly time series).
Consequently, the estimated TPDM ($\hat{\sigma}(1)=0.49$) is not as strong as when focused only on the upper tail ($\hat{\sigma}(1)=0.71$).
The last model for comparison is a (standard) linear regularly-varying time series restricted to take values in $\mathbb{R}^+$ by imposing that the parameter estimates be positive. 
This model was fitted to the same transformed anomalies as our model, and shares the same estimated TPDF. 
As our transformed-linear model resulted in a negative $\hat{\theta}$, restricting the ARMA coefficients to be positive results in $\hat{\theta}=0$ and we essentially get an AR(1) fit. 

To explore model performance in capturing tail dependence, we calculate some tail summary statistics, and compare them across the above discussed models. Table \ref{table_run} gives average length of run above higher quantiles for the actual time series data, the synthetic ARMA(1,1) time series generated from our transformed-linear regularly-varying model, and synthetic ARMA(1,1) time series generated from the three alternative models. Table \ref{table_sum} gives the higher quantiles for sum of three consecutive time series terms. In both tables, the transformed linear model performs best, and seems to produce reasonable estimates of these tail quantities. Likely due to its asymptotic independence, or perhaps due to the fact that it was fit to the entire data set rather than focusing on extreme behavior, the non-extreme Gaussian model underestimates both quantities. 
The linear regularly-varying model on $\mathbb{R}$, with regular variation in both directions, also exhibits a lower dependence in the upper tail because of the symmetry in the definition of the TPDF.
This provides evidence for restricting the model to the positive orthant when interest is only in the upper tail. The linear regularly-varying model on the positive orthant, with coefficients restricted to be positive, overestimates the dependence in the upper tail as it does not allow for a negative MA coefficient. Hence, the transformed-linear model provides more flexibility than this restricted model.

\begin{table}[h]
\fontsize{10}{10}\selectfont
 \begin{center}
    \caption{Average length of run above a threshold}
    \label{table_run}
        \begin{tabular}{lccccc}
        \hline
        Threshold & \multicolumn{5}{c}{Length of run above the quantile} \\ \cline{2-6} 
                             quantile   & \multicolumn{1}{c}{Actual} & \multicolumn{1}{c}{Trans-Lin Reg-Var}   & \multicolumn{1}{c}{Gaussian} & \multicolumn{1}{c}{Lin Reg-Var}  & \multicolumn{1}{c}{Lin Reg-Var} \\ 
                                 &  & \multicolumn{1}{c}{in $\mathbb{R}^+$}   &  & \multicolumn{1}{c}{in $\mathbb{R}$}  & \multicolumn{1}{c}{ in $\mathbb{R}^+$} \\ \hline
        0.95 & \multicolumn{1}{c}{2.43 (2.89)} & \multicolumn{1}{c}{2.60 (4.47)}  & \multicolumn{1}{c}{1.48 (1.09)} & \multicolumn{1}{c}{1.56 (2.19)} & \multicolumn{1}{c}{5.41 (5.69)} \\
        0.98 & \multicolumn{1}{c}{2.35 (2.59)} & \multicolumn{1}{c}{2.60 (4.42)}  & \multicolumn{1}{c}{1.34 (0.81)} & \multicolumn{1}{c}{1.52 (2.44)} & \multicolumn{1}{c}{5.74 (5.43)}\\ 
        0.99 & \multicolumn{1}{c}{2.10 (2.18)} & \multicolumn{1}{c}{2.66 (4.29)}  & \multicolumn{1}{c}{1.27 (0.67)} & \multicolumn{1}{c}{1.59 (2.73)} & \multicolumn{1}{c}{5.93 (5.38)}\\ 
        0.995 & \multicolumn{1}{c}{1.77 (1.82)} & \multicolumn{1}{c}{2.61 (4.22)}  & \multicolumn{1}{c}{1.21 (0.54)} & \multicolumn{1}{c}{1.75 (3.08)} & \multicolumn{1}{c}{5.83 (5.03)}\\ 
        0.999 & \multicolumn{1}{c}{1.40 (0.82)} & \multicolumn{1}{c}{2.21 (3.20)}  & \multicolumn{1}{c}{1.08 (0.28)} & \multicolumn{1}{c}{2.04 (2.84)} & \multicolumn{1}{c}{5.78 (4.48)}\\ \hline
        \end{tabular}
 \end{center}
\end{table}

\begin{table}[h]
 \fontsize{10}{10}\selectfont
 \begin{center}
    \caption{Quantiles for sum of three consecutive terms}
    \label{table_sum}
        \begin{tabular}{lccccc}
        \hline
        Quantile & Actual & Trans-Lin Reg-Var &   Gaussian & Lin Reg-Var & Lin Reg-Var \\
         &  & in $\mathbb{R}^+$  &  & in $\mathbb{R}$ & in $\mathbb{R}^+$ \\\hline
        0.95 & 8.28 & 8.45  & 7.93 & 7.45 & 8.92 \\
        0.98 & 13.09 & 12.91 & 11.62 & 11.01 & 13.79 \\ 
        0.99 & 16.79 & 16.72 & 14.34 & 14.08 & 17.74 \\ 
        0.995 & 20.29 & 20.46 & 17.02 & 18.00 & 21.47 \\ 
        0.999 & 27.17 & 28.09 & 23.00 & 27.22 & 29.24 \\ \hline
        \end{tabular}
 \end{center}
\end{table}


\section{Discussion}

This paper constructs straightforward, flexible, and interpretable time series models by applying transformed-linear operations to regularly-varying random variables.
Unlike other linear time series models, our time series models only take positive values, allowing one to focus modeling and inference entirely on the upper tail.
As is common in time series, we characterize dependence between pairs of elements, and the TPDF has properties analogous to those of the ACVF. 
Our notion of weak tail stationarity helps relax the assumption of iid noise terms and more importantly allows characterization of the time series' upper tail via the TPDF.
Our transformed-linear time series models have similar interpretations to and share some properties of their non-extreme ARMA analogues.
Application to the Santa Ana hourly windspeed time series shows that our models appropriately capture extremal dependence. 
Fitting the MA(1), AR(1), and ARMA(1,1) transformed-linear time series models requires only minutes.

Our fitted model's run length estimate (Table \ref{table_run}) seems to exhibit ``threshold stability," common to asymptotically dependent models.
Recent work, mainly in spatial extremes, has aimed to develop models with more nuanced handling of tail dependence (e.g., \citet{huseretal2018}, \citet{wadsworthtawn2019}, \citet{boppetal2020}).
There are likely time series analyses where similar models would prove beneficial; but these models and their estimation procedures are quite complex. We believe that there is value in simple models.

A current challenge in extremes is tail dependence estimation, which was evident when estimating the TPDF, particularly in cases of weak tail dependence.
We are confident that tail dependence estimation methods will continue to improve, aiding in TPDF estimation.

The fact that the tranformed linear MA($\infty$) class of models can be linked to a Hilbert space opens avenues for further exploration.
In future work we aim to investigate method-of-moments estimation procedures analogous to traditional time series methods such as Yule-Walker or the innovations algorithm. 
We would also like to investigate forecasting methods for transformed-ARMA models.

\section*{Acknowledgement}
Nehali Mhatre and Daniel Cooley were both partially supported by National Science Foundation award DMS-1811657.

\section*{Supplementary Material}
\label{SM}
Supplementary Material available online (\href{https://www.stat.colostate.edu/~cooleyd/TransLinTS/}{\textit{link}}) includes a proof that the space $\mathbb{V}$ of infinite transformed-linear combinations of regularly-varying random variables with absolutely summable coefficients is an inner product space and is isomorphic to the vector space of absolutely summable sequences $\ell^1$.
The Supplementary Material also demonstrates through simulation, the bias reduction in TPDF estimation by subtracting off the mean of the marginally transformed Fr\'echet time series.
The windspeed anomalies data and the R code to estimate TPDF for the data and to fit transformed-linear regularly-varying time series models discussed in Section 4 are also available at the \href{https://www.stat.colostate.edu/~cooleyd/TransLinTS/}{link}.


\appendix

\section{Appendix}
\label{appn}

\subsection{Proofs of propositions}
\label{a1}
\noindent {\underline{Proof of Proposition~\ref{prop_1}:}}  By the relation between limiting measure and angular measure, we have that
$\nu_{\mathbf{X}_{(l)}}\{A_{(l)}(r_0, B_{l-1})\} = r_0^{-\alpha} H_{\mathbf{X}_{(l)}} (B_{l-1}).$
Consider the limiting measure $\nu_{\mathbf{X}}\{A^*_{(l)}(r_0, B_{l-1})\}$.
\begin{align}
	\nu_{\mathbf{X}}\{A^*_{(l)}(r_0, B_{l-1})\} & = \int_{\mathbf{w} \in \mathbb{S}^+_{p-1} : \| \mathbf{w}_{(l)} \|^{-1} \mathbf{w}_{(l)} \in B_{l-1}} \int_{r = r_0/\|\mathbf{w}_{(l)}\|}^\infty \alpha r^{-(\alpha + 1)} \text{d}r\text{d}H_{\mathbf{X}}(\mathbf{w}) \nonumber \\
    & = r_0^{-\alpha} \int_{\mathbf{w} \in \mathbb{S}^+_{p-1} : \| \mathbf{w}_{(l)} \|^{-1} \mathbf{w}_{(l)} \in B_{l-1}} \| \mathbf{w}_{(l)} \|^{\alpha} \text{d}H_{\mathbf{X}}(\mathbf{w}). \nonumber
\end{align}
By consistency, $\nu_{\mathbf{X}}\{A^*_{(l)}(r_0, B_{l-1})\} = \nu_{\mathbf{X}_{(l)}}\{A_{(l)}(r_0, B_{l-1})\},$ and (\ref{prop1}) follows.\smallskip

\noindent {\underline{Proof of Proposition~\ref{prop_2}:}}  By construction, $v_i = x_i / \|(x_i,x_j)\|$ and $w_i = x_i / \|\mathbf{x}\|$.
Therefore, 
\begin{eqnarray}
v_i = w_i \frac{\|\mathbf{x}\|}{ \|(x_i,x_j)\|} = w_i \frac{\sqrt{x_1^2+ \cdots + x_p^2}}{ \sqrt{x_i^2+x_j^2}} = w_i \frac{\sqrt{\frac{x_1^2}{\|\mathbf{x}\|^2}+ \cdots + \frac{x_p^2}{\|\mathbf{x}\|^2}}}{ \sqrt{\frac{x_i^2}{\|\mathbf{x}\|^2}+\frac{x_j^2}{\|\mathbf{x}\|^2}}} = w_i \frac{\sqrt{w_1^2 + \cdots + w_p^2}}{\sqrt{w_i^2+ w_j^2}} = \frac{w_i}{\|(w_i,w_j)\|}.\nonumber
\end{eqnarray}
Similarly, $v_j = w_j / \|(w_i,w_j)\|$.
Therefore, by Proposition 2.1, for $l=2$ and $\alpha=2$, we have,
\begin{align}
	\sigma(X_i, X_j) & = \int_{\mathbf{v} \in {\mathbb{S}}^+_1} v_iv_j \text{d}H_{(X_i, X_j)}(\mathbf{v}) = \int_{\mathbf{w} \in \mathbb{S}^+_{p-1}} \frac{w_i}{\|(w_i,w_j)\|} \frac{w_j}{\|(w_i,w_j)\|} \| (w_i, w_j) \|^{2} \text{d}H_{\mathbf{X}}(\mathbf{w}) \nonumber \\
	& = \int_{\mathbf{w} \in \mathbb{S}^+_{p-1}} w_i w_j \text{d}H_{\mathbf{X}}(\mathbf{w}). \nonumber
\end{align}

\subsection{Derivation of equation (\ref{eq52}) in the manuscript}
\label{a2}

\noindent{Applying $ \pi(B) =  \sum_{j=0}^\infty \phi^j B^j$ to $\{\pi(B) \phi(B)\} \circ X_t = \pi(B) \circ Z_t$, we get,}
\begin{align} \label{app:a}
  \left\{\sum_{j=0}^\infty \phi^j B^j (1-\phi B)\right\} \circ X_t = \left(\sum_{j=0}^\infty \phi^j B^j\right) \circ Z_t.
\end{align}
Let us first consider the left hand side of (\ref{app:a}):
\begin{align} \label{app:b}
 \left\{\sum_{j=0}^\infty \phi^j B^j (1-\phi B)\right\} \circ X_t & = f \left\{\sum_{j=0}^\infty \phi^j B^j (1-\phi B) f^{-1}(X_t)  \right\} \nonumber \\
 & = f \left\{\sum_{j=0}^\infty \phi^j B^j f^{-1}(X_t) - \sum_{j=0}^\infty \phi^{j+1} B^{j+1} f^{-1}(X_t)  \right\} \nonumber \\
 & = f \{f^{-1}(X_t) \} = X_t.
\end{align}
Let us now consider the right hand side of (\ref{app:a}):
\begin{align} \label{app:c}
 \left(\sum_{j=0}^\infty \phi^j B^j\right) \circ Z_t & = f \left\{ \sum_{j=0}^\infty \phi^j B^j f^{-1}(Z_t) \right\} = f \left\{ \sum_{j=0}^\infty \phi^j f^{-1}(Z_{t-j}) \right\} = \bigoplus_{j=0}^\infty \phi^j \circ Z_{t-j}.
\end{align}
Putting together (\ref{app:b}) and (\ref{app:c}), $X_t = \bigoplus_{j=0}^\infty \phi^j \circ Z_{t-j}.$

\subsection{Derivation of equation (\ref{eq69}) in the manuscript}
\label{a3}

\noindent{Expanding $\psi(B)$, we get,}
\begin{align} \label{app:69}
    \psi(B) & = \pi(B) \theta(B) = \left(\sum^{\infty}_{j=0}\phi^jB^j\right) (1+\theta B) = (1+\phi B + \phi^2 B^2 + \phi^3 B^3 + \cdots ) (1+\theta B) \nonumber \\
        & = 1+ (\phi+\theta)B + (\phi+\theta)\phi B^2 + (\phi+\theta)\phi^2 B^3 + \cdots = \sum^{\infty}_{j=0}\psi_jB^j,
\end{align}
where $\psi_0 = 1$ and $\psi_j = (\phi+\theta)\phi^{j-1}$ for $j\ge1$.

\noindent{Substituting (\ref{app:69}) in equation (\ref{eq68}) of the manuscript, we get,}
\begin{align} 
    X_t & = \left(\sum^{\infty}_{j=0}\psi_jB^j\right) \circ Z_t \nonumber \\
        & = \{1+ (\phi+\theta)B + (\phi+\theta)\phi B^2 + (\phi+\theta)\phi^2 B^3 + \cdots \} \circ Z_t \nonumber \\
        & = f[\{1+ (\phi+\theta)B + (\phi+\theta)\phi B^2 + (\phi+\theta)\phi^2 B^3 + \cdots \} f^{-1}(Z_t)] \nonumber \\
        & = f\{f^{-1}(Z_t) + (\phi+\theta)Bf^{-1}(Z_t) + (\phi+\theta)\phi B^2f^{-1}(Z_t) + (\phi+\theta)\phi^2 B^3f^{-1}(Z_t) + \cdots  \} \nonumber \\
        & = f\{f^{-1}(Z_t) + (\phi+\theta)f^{-1}(Z_{t-1}) + (\phi+\theta)\phi f^{-1}(Z_{t-2}) + (\phi+\theta)\phi^2 f^{-1}(Z_{t-3}) + \cdots  \} \nonumber \\
        & = Z_t \oplus (\phi+\theta) \circ \left(\bigoplus^{\infty}_{j=1}\phi^{j-1} \circ Z_{t-j}\right) \nonumber
\end{align}

\subsection{Derivation of equation (\ref{eq71}) in the manuscript}
\label{a4}

\noindent{Letting $\zeta(B) = -\sum^{\infty}_{j=1}\phi^{-j}B^{-j}$, and applying $\zeta(B)$ to both sides of equation (30) in the manuscript, we get}
\begin{align}
    X_t & = \left\{\left(-\sum^{\infty}_{j=1}\phi^{-j}B^{-j}\right)(1+\theta B)\right\} \circ Z_t \nonumber \\
        & = \left\{(-\phi^{-1}B^{-1} - \phi^{-2}B^{-2} - \phi^{-3}B^{-3} - \cdots )(1+\theta B)\right\} \circ Z_t \nonumber \\
        & = \left\{-\phi^{-1}\theta - (\phi+\theta)\phi^{-2}B^{-1} - (\phi+\theta)\phi^{-3} B^{-2} - \cdots \right\} \circ Z_t \nonumber \\
        & = f\left[\left\{-\phi^{-1}\theta - (\phi+\theta)\phi^{-2}B^{-1} - (\phi+\theta)\phi^{-3} B^{-2} - \cdots \right\} f^{-1}(Z_t)\right] \nonumber \\
        & = f\left\{-\phi^{-1}\theta f^{-1}(Z_t) - (\phi+\theta)\phi^{-2}B^{-1} f^{-1}(Z_t) - (\phi+\theta)\phi^{-3} B^{-2} f^{-1}(Z_t) - \cdots  \right\} \nonumber \\
        & = f\left\{-\phi^{-1}\theta f^{-1}(Z_t) - (\phi+\theta)\phi^{-2}f^{-1}(Z_{t+1}) - (\phi+\theta)\phi^{-3}  f^{-1}(Z_{t+2}) - \cdots  \right\} \nonumber \\
        & = -\phi^{-1}\theta Z_t \oplus -(\phi+\theta) \circ \left(\bigoplus^{\infty}_{j=1}\phi^{-j-1} \circ Z_{t+j}\right). \nonumber
\end{align}

\subsection{Derivation of the TPDF expression for a transformed-linear regularly-varying ARMA(1,1) time series model}
\label{a5}

\noindent{The general form for the TPDF of a transformed-linear regularly-varying ARMA(1,1) time series is given as,}
$$\sigma(h) = \sum^{\infty}_{j=0} \psi_j^{(0)} \psi_{j+h}^{(0)},$$
where $\psi_0=1$, $\psi_j=(\phi+\theta)\phi^{j-1}$ for $j>0$, $\psi_j^{(0)}= \text{max}(\psi_j,0)$, and $\sum^{\infty}_{j=0}|\psi_j|<\infty$.
We consider different cases of $\phi$ and $\theta$ below.

\noindent{Case} 1: $\phi>0$ and $\phi+\theta>0$, so that $\psi_0=1$ and $\psi_j>0$ for all $j$.

\noindent{For} $h>0$,
\begin{align} \label{arma11_1}
    \sigma(0) = \sum^{\infty}_{j=0} \psi_j \psi_{j+h} & = \psi_0\psi_h + \sum^{\infty}_{j=1}(\phi+\theta)^2\phi^{j-1}\phi^{j+h-1} = (\phi+\theta)\phi^{h-1} + (\phi+\theta)^2\sum^{\infty}_{k=0}\phi^{2k+h} \nonumber \\
        & = (\phi+\theta)\phi^{h-1} + (\phi+\theta)^2\phi^h\sum^{\infty}_{k=0}\phi^{2k} = (\phi+\theta)\phi^{h-1} + \frac{(\phi+\theta)^2\phi^h}{1-\phi^2}  \nonumber \\
        & = (\phi+\theta)\phi^h \left\{ \frac{1}{\theta}+ \frac{(\phi+\theta)}{1-\phi^2} \right\} = \frac{(\phi+\theta)\phi^h (1+\phi\theta)}{1-\phi^2}.
\end{align}
For $h=0$,
\begin{align} \label{arma11_2}
    \sigma(h) = \sum^{\infty}_{j=0} \psi_j \psi_{j} & = \psi_0\psi_0 + \sum^{\infty}_{j=1}(\phi+\theta)^2\phi^{j-1}\phi^{j-1} = 1 + (\phi+\theta)^2\sum^{\infty}_{k=0}\phi^{2k} \nonumber \\
        & = 1 + \frac{(\phi+\theta)^2}{1-\phi^2}  = \frac{1+2\phi\theta+\theta^2}{1-\phi^2}.
\end{align}

\noindent{Case} 2: $\phi>0$ and $\phi+\theta<0$, so that $\psi_0=1$ and $\psi_j<0$ for all $j>0$.
\begin{align} \label{arma11_3}
    \sigma(h) & = \begin{cases}
        1, & \text{ if } h=0,\\
        0, & \text{ if } h>0.
    \end{cases}
\end{align}

\noindent{Case} 3: $\phi<0$ and $\phi+\theta>0$, so that $\psi_0=1$, $\psi_j<0$ if $j$ is even, and $\psi_j>0$ if $j$ is odd.

\noindent{For} $h>0$,
\begin{align} \label{arma11_4}
    \sigma(h) & = \sum^{\infty}_{j=0} \psi_j^{(0)} \psi_{j+h}^{(0)}  = \psi_h^{(0)} + \psi_1\psi_{h+1}^{(0)} + \psi_3\psi_{h+3}^{(0)} + \psi_5\psi_{h+5}^{(0)} + \cdots, \text{ as } \psi_j<0 \text{ if } j \text{ is even,} \nonumber \\
     & = \begin{cases}
         0 + \psi_1\psi_{h+1} + \psi_3\psi_{h+3} + \psi_5\psi_{h+5} + \cdots, & \text{ if } h \text{ is even,} \\
         \psi_h, &  \text{ if } h \text{ is odd,}
     \end{cases} \nonumber \\
          & = \begin{cases}
         (\phi+\theta)\phi^0 (\phi+\theta)\phi^{h} + (\phi+\theta)\phi^2 (\phi+\theta)\phi^{h+2} + (\phi+\theta)\phi^4 (\phi+\theta)\phi^{h+4} + \cdots, & \text{ if } h \text{ is even,} \\
         \psi_h, & \text{ if } h \text{ is odd,}
     \end{cases} \nonumber \\
               & = \begin{cases}
         (\phi+\theta)^2\phi^{h} + (\phi+\theta)^2\phi^{h+4} + (\phi+\theta)^2\phi^{h+8} + \cdots, & \text{ if } h \text{ is even,} \\
         (\phi+\theta)\phi^{h-1}, & \text{ if } h \text{ is odd,}
     \end{cases} \nonumber \\
                    & = \begin{cases}
         (\phi+\theta)^2\phi^{h} \left(1  + \phi^{4} + \phi^{8} + \cdots \right), & \text{ if } h \text{ is even,} \\
         (\phi+\theta)\phi^{h-1}, & \text{ if } h \text{ is odd,}
     \end{cases} \nonumber \\
                    & = \begin{cases}
         \frac{(\phi+\theta)^2\phi^{h}}{1-\phi^4}, & \text{ if } h \text{ is even,} \\
         (\phi+\theta)\phi^{h-1}, & \text{ if } h \text{ is odd.}
     \end{cases}
\end{align}

\noindent{For} $h=0$,
\begin{align} \label{arma11_5}
    \sigma(h) & = \sum^{\infty}_{j=0} \psi_j^{(0)} \psi_{j}^{(0)} = 1 + \psi_1^2 + \psi_3^2 + \psi_5^2 + \cdots \nonumber \\
    & = 1 + (\phi+\theta)^2\left(1+\phi^4+\phi^8  + \cdots \right) = 1 + \frac{(\phi+\theta)^2}{1-\phi^4} = \frac{1-\phi^4+(\phi+\theta)^2}{1-\phi^4}
\end{align}

\noindent{Case} 4: $\phi<0$ and $\phi+\theta<0$, so that $\psi_0=1$, $\psi_j>0$ if $j$ is even, and $\psi_j<0$ if $j$ is odd.

\noindent{For} $h>0$,
\begin{align} \label{arma11_6}
    \sigma(h) & = \sum^{\infty}_{j=0} \psi_j^{(0)} \psi_{j+h}^{(0)}  = \psi_h^{(0)} + \psi_2\psi_{h+2}^{(0)} + \psi_4\psi_{h+4}^{(0)} + \psi_6\psi_{h+6}^{(0)} + \cdots, \text{ as } \psi_j<0 \text{ if } j \text{ is odd,} \nonumber \\
     & = \begin{cases}
         \psi_h + \psi_2\psi_{h+2} + \psi_4\psi_{h+4} + \psi_6\psi_{h+6} + \cdots, & \text{ if } h \text{ is even,} \\
         0, &  \text{ if } h \text{ is odd,}
     \end{cases} \nonumber \\
          & = \begin{cases}
         (\phi+\theta)\phi^{h-1} + (\phi+\theta)\phi (\phi+\theta)\phi^{h+1} + (\phi+\theta)\phi^3 (\phi+\theta)\phi^{h+3} + \cdots, & \text{ if } h \text{ is even,} \\
         0, &  \text{ if } h \text{ is odd,}
     \end{cases} \nonumber \\
               & = \begin{cases}
         (\phi+\theta)\phi^{h-1} + (\phi+\theta)^2\phi^{h+2} \left(1 + \phi^4 + \phi^8 +  \cdots \right), &  \text{ if } h \text{ is even,} \\
         0, &  \text{ if } h \text{ is odd,}
     \end{cases} \nonumber \\
                & = \begin{cases}
         (\phi+\theta)\phi^{h-1} + \frac{(\phi+\theta)^2\phi^{h+2}}{1-\phi^4}, &  \text{ if } h \text{ is even,} \\
         0, &  \text{ if } h \text{ is odd.}
     \end{cases}
\end{align}

\noindent{For} $h=0$,
\begin{align} \label{arma11_7}
    \sigma(h) & = \sum^{\infty}_{j=0} \psi_j^{(0)} \psi_{j}^{(0)} = 1 + \psi_2^2 + \psi_4^2 + \psi_6^2 + \cdots \nonumber \\
    & = 1 + (\phi+\theta)^2\phi^2\left(1+\phi^4+\phi^8  + \cdots \right) = 1 + \frac{(\phi+\theta)^2\phi^2}{1-\phi^4} = \frac{1+\phi^2\theta^2+2\phi^3\theta}{1-\phi^4}
\end{align}
Putting together (\ref{arma11_1}) to (\ref{arma11_7}), for ARMA(1,1) with tail ratio 1, i.e. $\sigma(0)=1$, the TPDF for lag $h>0$ is given by,
\begin{align}
    \sigma(h) & =
                \begin{cases}
                \frac{(\phi+\theta) \phi^h (1+\phi\theta)}{1+2\phi\theta + \theta^2} & \text{if } \phi > 0, \phi+\theta>0 \\
                0 & \text{if } \phi > 0, \phi+\theta<0 \\
                \frac{(\phi+\theta)^2\phi^h}{1-\phi^4+(\phi+\theta)^2} & \text{if } \phi < 0, \phi+\theta>0, h \text{ is even} \\
                \frac{(\phi+\theta)\phi^{h-1}(1-\phi^4)}{1-\phi^4+(\phi+\theta)^2} & \text{if } \phi < 0, \phi+\theta>0, h \text{ is odd} \\
                \frac{(\phi+\theta)\phi^{h-1}(1+\theta\phi^3)}{1+\phi^2\theta^2+2\phi^3\theta} & \text{if } \phi < 0, \phi+\theta<0, h \text{ is even} \\
                0 & \text{if } \phi < 0, \phi+\theta<0, h \text{ is odd}, \nonumber
                \end{cases} \nonumber
\end{align}


\begin{thebibliography}{7}
\expandafter\ifx\csname natexlab\endcsname\relax\def\natexlab#1{#1}\fi

  
  
\bibitem[{Bopp et~al.(2020)Bopp, Shaby \& Huser}]{boppetal2020}
Bopp, G.~P., Shaby, B.~A. \& Huser, R. (2020).
\newblock A hierarchical max-infinitely divisible spatial model for extreme precipitation.
\newblock \textit{Journal of the American Statistical Association.} pp.~1--14.

\bibitem[{Brockwell \& Davis(1991)}]{brockwelldavis1991}
Brockwell, P.~J. \& Davis, R.~A. (1991).
\newblock \textit{Time series: theory and methods.}
\newblock Springer, New York.

\bibitem[{Brockwell \& Davis(2002)}]{brockwelldavis2002}
Brockwell, P.~J. \& Davis, R.~A. (2002).
\newblock \textit{Introduction to time series and forecasting.} Second edition.
\newblock Springer, New York.


\bibitem[{Coles(2001)}]{coles2001}
Coles, S. (2001).
\newblock \textit{An introduction to statistical modeling of extreme values.}
\newblock Springer-Verlag, London.

\bibitem[{Cooley \& Thibaud(2019)}]{cooleythibaud2019}
Cooley, D. \& Thibaud, E. (2019).
\newblock Decompositions of dependence for high-dimensional extremes.
\newblock \textit{Biometrika.} {\bf 106}(3),~587--604.

  
\bibitem[{Davis \& Mikosch(2009)}]{davis2009}
Davis, R.~A. \& Mikosch, T. (2009).
\newblock The extremogram: A correlogram for extreme events.
\newblock \textit{Bernoulli.} {\bf 15}(4),~977--1009.

\bibitem[{Davis \& Resnick(1989)}]{davisresnick1989}
Davis, R.~A. \& Resnick, S.~I. (1989).
\newblock Basic properties and prediction of max-arma processes.
\newblock \textit{Advances in Applied Probability.} {\bf 21}(4),~781--803.


  
\bibitem[{Dunn(2019)}]{dunn2019}
Dunn, R. J.~H. (2019).
\newblock Hadisd version 3: Monthly updates. hadley centre tech, Technical report, Note 103, 10 pp., www.metoffice.gov.uk/research/library-and-archive

\bibitem[{Embrechts et~al.(1997)Embrechts, Mikosch \& Kl{\"u}ppelberg}]{embrechtsetal1997}
Embrechts, P., Mikosch, T. \& Kl{\"u}ppelberg, C. (1997).
\newblock \textit{Modelling Extremal Events: For Insurance and Finance.}
\newblock Springer-Verlag, Berlin, Heidelberg.

\bibitem[{Huser et~al.(2016)Huser, Davison \& Genton}]{huseretal2016}
Huser, R., Davison, A.~C. \& Genton, M.~G. (2016).
\newblock Likelihood estimators for multivariate extremes.
\newblock \textit{Extremes.} {\bf 19}(1),~79--103.

\bibitem[{Huser et~al.(2018)Huser, Opitz \& Thibaud}]{huseretal2018}
Huser, R., Opitz, T. \& Thibaud, E. (2018).
\newblock Max-infinitely divisible models and inference for spatial extremes.
\newblock \textit{arXiv preprint.}
\newblock arXiv:1801.02946.

\bibitem[{Kim \& Kokoszka(2020)}]{kimkokoszka2020}
Kim, M. \& Kokoszka, P. (2020).
\newblock Extremal dependence measure for functional data.
\newblock \textit{Preprint.}

\bibitem[{Kulik \& Soulier(2020)}]{kuliksoulier2020}
Kulik, R. \& Soulier, P. (2020).
\newblock \textit{Heavy-tailed time series.}
\newblock Springer Series in Operations Research and Financial Engineering.
\newblock Springer-Verlag, New York.

\bibitem[{Larsson \& Resnick(2012)}]{larssonresnick2012}
Larsson, M. \& Resnick, S.~I. (2012).
\newblock Extremal dependence measure and extremogram: the regularly varying case.
\newblock \textit{Extremes.} {\bf 15}(2),~231--256.


\bibitem[{Resnick(1987)}]{resnick1987}
Resnick, S.~I. (1987).
\newblock \textit{Extreme Values, Regular Variation and Point Processes.}
\newblock Springer Series in Operations Research and Financial Engineering.
\newblock Springer-Verlag, New York.

\bibitem[{Resnick(2007)}]{resnick2007}
Resnick, S.~I. (2007).
\newblock \textit{Heavy-tail phenomena: probabilistic and statistical modeling.}
\newblock Springer-Verlag, New York.

\bibitem[{Sklar(1959)}]{sklar1959}
Sklar, M. (1959).
\newblock Fonctions de R{\'e}partition {\`A} N Dimensions Et Leurs Marges.
\newblock \textit{Publications de l`Institut Statistique de l`Universit{\'e} de Paris.} 8, 229-231.
  
\bibitem[{Smith et~al.(1997)}]{smithetal}
Smith, R.~L., Tawn, J. \& Coles, S. (1997).
\newblock Markov chain models for threshold exceedances.
\newblock \textit{Biometrika.}, 84, 249-268.
  
\bibitem[{Strokorb \& Schlather(2015)}]{strokorbschlather2015}
Strokorb, K. \& Schlather, M. (2015).
\newblock An exceptional max-stable process fully parameterized by its extremal coefficients.
\newblock \textit{Bernoulli.} {\bf 21}(1),~276--302.

\bibitem[{Wadsworth \& Tawn(2019)}]{wadsworthtawn2019}
Wadsworth, J.~L. \& Tawn, J. (2019).
\newblock Higher-dimensional spatial extremes via single-site conditioning.
\newblock \textit{arXiv preprint.}
\newblock arXiv:1912.06560.

\end{thebibliography}

\end{document}